\documentclass[a4paper]{aa}
  \usepackage{amsmath,ifpdf,txfonts,natbib,graphicx}
  \usepackage[english]{babel} 

\newcommand{\eqab}{\begin{eqnarray}} \newcommand{\eqae}{\end{eqnarray}}
\newcommand{\eqb}{\begin{equation}} \newcommand{\eqe}{\end{equation}}

\newcommand{\Cr}{\mathrm{C}}
\newcommand{\cm}{\mathrm{cm}}
\newcommand{\dr}{\mathrm{d}}
\newcommand{\er}{\mathrm{e}}

\newcommand{\emr}{\mathrm{em}}
\newcommand{\Hr}{\mathrm{H}}
\newcommand{\He}{\mathrm{He}}
\newcommand{\K}{\mathrm{K}}
\newcommand{\keV}{\mathrm{KeV}}
\newcommand{\Or}{\mathrm{O}}
\newcommand{\pe}{\mathrm{pe}}
\newcommand{\yr}{\mathrm{yr}^{-1}}

\newcommand{\Bb}{\mathbf{B}}
\newcommand{\Jb}{\mathbf{J}}
\newcommand{\vb}{\mathbf{v}}
\newcommand{\kms}{\mathrm{km~s}^{-1}}
\newcommand{\hs}{\mathrm{hs}}
\newcommand{\Macc}{\dot{M}_\mathrm{acc}}
\newcommand{\Msun}{M_\odot}
\newcommand{\Mstar}{M_\star}
\newcommand{\Tstar}{T_\star}
\newcommand{\Rstar}{R_\star}
\newcommand{\Rsun}{R_\odot}
\newcommand{\Rsub}{R_\mathrm{sub}}
\newcommand{\Rmin}{R_{\rm min}}
\newcommand{\Rin}{R_{\rm in}}
\newcommand{\rin}{r_{\rm in}}
\newcommand{\rout}{r_{\rm out}}
\newcommand{\rA}{r_{\rm A}}
\newcommand{\vin}{v_{\rm in}}
\newcommand{\vp}{V_{\rm p}}

  \ifpdf
    \usepackage[pdftex,unicode,pdfauthor=despo,colorlinks,breaklinks=true]
    {hyperref}
  \else
    \usepackage[dvips,unicode,pdfauthor=despo,colorlinks,breaklinks=true]
    {hyperref}
  \fi 

  \titlerunning{Molecules in protostellar MHD disk winds. I. Model}
  \authorrunning{Panoglou et al.}

  \title{Molecule survival in magnetized protostellar disk winds}
  \subtitle{I. Chemical model and first results}
  \author{
          D.~Panoglou\inst{1,2}
          \thanks{
          Present address: I. Physikalisches Institut,
          Universit\"at zu K\"oln, Z\"ulpicher Stra{\ss}e 77,
          50937 K\"oln, Germany
          }
     \and S.~Cabrit\inst{2}
     \and G.~Pineau des For\^ets\inst{3,2}
     \and P.~J.~V.~Garcia\inst{4,5}        
     \and J.~Ferreira\inst{5}        
     \and F.~Casse\inst{6}
     }

\date{Received 10 July 2009 / Accepted 14 November 2011}
\offprints{\tt panoglou@ph1.uni-koeln.de}

\institute{
        Centro de Astrof\'isica, Universidade do Porto,
        4150-752 Porto, Portugal 
   \and LERMA, Observatoire de Paris, ENS, UPMC, UCP, 
        CNRS, 61 Avenue de l'Observatoire, 75014 Paris, France
   \and IAS, UMR 8617 du CNRS, Universit\'e de Paris-Sud,
        91405 Orsay, France
   \and Universidade do Porto, Faculdade de Engenharia,
        Laborat\'orio SIM Unidade FCT \# 4006, Portugal
   \and Institut de Plan\'etologie et d'Astrophysique de Grenoble,
        UMR 5521 du CNRS, 38041 Grenoble Cedex, France
   \and Laboratoire Astroparticule \& Cosmologie, Universit\'e Paris 7,
        UMR 7164 du CNRS, 75205 Paris Cedex 13, France
        }

\begin{document}
\abstract
{Molecular counterparts to atomic jets have recently been detected within 1000 AU of young stars at early evolutionary stages. Reproducing these counterparts is an important new challenge for proposed ejection models.}
{We explore whether molecules may survive in the magneto-hydrodynamic (MHD) disk wind solution currently invoked to reproduce the kinematics and tentative rotation signatures of atomic jets in T Tauri stars.}
{The coupled ionization, chemical, and thermal evolution along dusty flow streamlines is computed for the prescribed MHD disk wind solution, using a method developed for magnetized shocks in the interstellar medium. Irradiation by (wind-attenuated) coronal X-rays and far-ultraviolet photons from accretion hot spots is included, with an approximate self-shielding of H$_2$ and CO. Disk accretion rates of $5\times10^{-6}$, $10^{-6}$ and $10^{-7}\Msun\yr$ are considered, representative of low-mass young protostars (so-called `Class~0'), evolved protostars (`Class~I') and very active T~Tauri stars (`Class~II') respectively. }
{The disk wind has an "onion-like" thermo-chemical structure, with streamlines launched from larger radii having lower temperature and ionisation, and higher H$_2$ abundance.The coupling between charged and neutral fluids is sufficient to eject molecules from the disk out to at least 9~AU. The launch radius beyond which most H$_2$ survives moves outward with evolutionary stage, from $\simeq0.2$~AU (sublimation radius) in the Class~0 disk wind, to $\simeq1$~AU in the Class~I, and $>1$~AU in the Class~II. In this molecular wind region, CO survives in the Class~0 but is significantly photodissociated in the Class~I/II. Balance between ambipolar heating and molecular cooling establishes a moderate asymptotic temperature $\simeq700-3000$~K, with cooler jets at earlier protostellar stages. As a result, endothermic formation of H$_2$O is efficient, with abundances up to $\simeq10^{-4}$, while CH$^+$ and SH$^+$ can reach $\geq10^{-6}$ in the hotter and more ionised Class~I/II winds.} 
{A centrifugal MHD disk wind launched from beyond $0.2-1$~AU can produce molecular jets/winds up to speeds $\simeq100~\kms$ in young low-mass stars ranging from Class~0 to active Class~II. The models predicts a high ratio of H$_2$ to CO and an increase of molecular launch radius, temperature, and flow width as the source evolves, in promising agreement with current observed trends. Calculations of synthetic maps and line profiles in H$_2$, CO and H$_2$O will allow detailed tests of the model against observations.}

\maketitle

\section{Introduction}
Powerful supersonic bipolar jets are ubiquitous in young accreting stars, suggesting that jets may play a key role in extracting excess angular momentum from the accretion flow. The high ejection efficiency, and the universal jet collimation properties independently of circumstellar envelope density, both strongly favor a magneto-hydrodynamical (MHD) ejection and collimation process (see e.g.~\citealt{KoPu00,shu00,cabrit-villard}). However, the main jet origin is still the subject of intense debate. Proposed options include a self-collimated, steady centrifugally-driven disk wind --- either from the inner disk edge near corotation ("X-wind") or from a more extended range of radii --- a pressure-driven stellar wind, and unsteady ejections by reconnexions in the sheared stellar magnetosphere, all three processes probably coexisting to some degree in young stars (see \citealt{Ferr06} and refs.~therein).

Of the wide variety of models published, stationary solutions for "X-winds" and extended MHD disk-winds are of particular interest as they provide synthetic predictions in atomic lines that reproduce several aspects of jet collimation and acceleration in microjets from T Tauri stars \citep{Shang98,SGSL02,Cab99,Garcia01b,Pes04,pyo-DG,pyo-HL-RW,Cabrit-IAU243}. Extended MHD disk winds have received further attention since the tentative detections of jet rotation signatures, that suggest an outer launch radius for the atomic jet%
  \footnote{We will use the term "atomic jet" rather than "optical jet" to
  refer to the (mostly neutral) fast jet component emitting in atomic/ionic
  lines, as infrared detections e.g.~in [\ion{Fe}{ii}] are increasingly
  frequent \citep[see][for a review]{Bally-PPV}. Atomic jets differ by their
  high speed and collimation from the "neutral atomic outflows" detected in
  \ion{H}{i} in a few young protostars, whose bulk mass is moving at
  $\leq20~\kms$ \citep{lizano88,svs13}.}
in T Tauri stars $\simeq0.2-3$~AU, well beyond the disk inner edge \citep{bac02,anderson03,Pes04,cof04,cof07}. Evidence for depletion of refractory species in atomic jets also suggests ejection from relatively large disk radii, outside the dust sublimation radius \citep{Nis05, Pod06, Pod09, Amb11}. If they remove most of the angular momentum from the accretion flow, such extended MHD disk winds would have a strong impact on the conditions for planet formation and migration in disks \citep{CombetFerr}. Therefore it is important to find further observational tests of their presence and radial extent.

An important new test for proposed jet engine(s) is whether they can explain the {\it molecular} jet/wind counterparts recently detected within 1000~AU of young stars. Molecular jets were first discovered emanating from the so-called `Class~0' protostars at the earliest stage of star formation, in CO, SiO, H$_2$, and H$_2$O masers \citep{Bach90, gui92,gg99,hh211-h2,claussen}. These narrow molecular jets are clearly distinct from the "standard" swept-up molecular outflows in that they are much faster, with deprojected speeds of $60-150~\kms$ as opposed to $\le20~\kms$ in the swept-up gas, much more highly collimated on-axis, and with specific chemical enhancements \citep{Bach90,gg99,tafalla09,tafalla10}. Their collimation scale, ejection/accretion ratio, and variation of physical parameters with velocity are strikingly similar to those of atomic T~Tauri jets, pointing to a similar origin \citep{Cab07,Lee07,Nis-SiO}. Counterparts in H$_2$, and more rarely in CO, were then identified towards more evolved infrared protostars of Class~I \citep{CR96,davis01,davis02,davis-AOsvs13,Bally-PPV} and optically visible T Tauri stars of Class~II \citep{tak-dgtau,tak-hltau,herczeg06,hh30-flow,beck08}. They consist of high-speed H$_2$ in the jet beam itself, and/or of slow wind/cavities at $\simeq30~\kms$ encompassing the fast ionic jet, on scales of a few 100~AU. The molecular emission tends to peak at lower velocities than atomic lines \citep{davis03}.

Three processes are generally invoked to produce molecular jets/winds in young stars: (i) entrainment of ambient molecular material, (ii) gas-phase reformation of H$_2$ in a dust-free atomic wind (e.g.~stellar or magnetospheric); (iii) ejection/reformation of molecules in a dusty wind from the disk surface.

The entrainment hypothesis is intuitively appealing, but appears quite challenging. Turbulent mixing of ambient gas along the walls of a supersonic jet beam predicts too low columns of H$_2$ \citep{taylor}. Shock entrainment in narrow bowshock wings along the jet is potentially more efficient, but would require that ambient molecular gas can quickly mix-in and refill the cocoon behind preceding bowshocks \citep{davis-hh111,raga93}. Swept-up molecular cavities driven into ambient gas by an atomic wide-angle wind have also been proposed in Class~II sources where molecular counterparts are broader than the atomic jet \citep[see e.g.][]{tak-hltau}. However, the nature of such a wide-angle component remains unclear; the fast $\simeq100~\kms$ and powerful wide-angle wind predicted by the "X-wind" model, propagating into a flattened envelope, would sweep up a cavity of width to length $>1$ at 200~AU from the source in less than 1000 years \citep{lee01,Shang06}. This is difficult to reconcile with the moderate width/length $<1$ of molecular "cavities" in Class~II sources of age $\simeq10^6$ yrs, unless the wide-angle wind is much slower/weaker (e.g.~a disk wind launched from large radii). In any case, an ambient swept-up cavity would not easily explain the narrow, high-velocity molecular jets seen in Class~0/I sources. The presence of blueshifted H$_2$ or CO up to $50-100~\kms$ within 10~AU of Class~II / Class~I sources (\citealt{herczeg06} and Herczeg, private communication) also seems difficult to explain with entrainment.

Alternatively, it has been argued that molecular counterparts may trace the primary ejected material itself \citep{CR96,tak-dgtau,hh30-flow,hirano}. Studying the formation/survival of H$_2$ in protostellar winds is a difficult problem as it requires the combination of state of the art MHD wind models with chemical networks and realistic radiation fields and heating/cooling terms, few authors having addressed it self-consistently in the past.

Reformation of H$_2$ in {\it dust-free stellar winds} by ion reactions has been the most extensively studied so far. Spherical models with prescribed temperature laws showed that H$_2$ forms and survives only at high mass-flux rate and if temperature drops rapidly below 2000~K \citep{Raw88,Glass89,Raw93}. \citet{Ruden90} solved in parallel for the ionisation, H$_2$ abundance, and thermal evolution, including ion-neutral "drag" heating. They found that less dense winds are hotter and more collisionally dissociated, so that the final H$_2$ abundance exceeds a few percents only at very high wind mass-fluxes $>10^{-5}\Msun~\yr$. Such a model does not seem able to explain the presence of detectable H$_2$ in low-luminosity Class~0/I/II jet sources, where total mass-flux rates are $1-3$ orders of magnitude smaller \citep{Lee07,antoniucci,HEG95}. Far-ultraviolet excess from accretion further reduces the predicted H$_2$ (and SiO) abundance at low wind mass-flux, even though CO is efficiently formed \citep{Glass91}. Alternatively, \citet{raga-h2form} show that H$_2$ abundances of up to 10\% could be reached behind dense internal shocks in a dust-free jet. However, the jet velocity variability amplitude should be less than $15~\kms$, which is quite restrictive. The effect of magnetic cushioning and of far-ultraviolet (FUV) photons remains to be investigated.

Here we investigate the third possibility, namely whether molecular jet counterparts could be tracing {\it dusty} disk winds. We focus on self-similar, centrifugally-driven MHD disk winds. Pioneering calculations by \citet{Safi93a} indicated that H$_2$ could survive collisional dissociation in the outer regions of such disk winds, beyond a minimum launch radius $\sim0.5-3$~AU depending on the adopted MHD solution. However, the rate of collisional dissociation may have been underestimated, as the ionisation fraction was obtained assuming atomic gas. The faster dissociative recombinations occuring in molecular gas would enhance ion-neutral drag, increasing the wind temperature and the dissociation rate. The calculations also did not consider H$_2$ destruction by stellar FUV photons, coronal X-rays, or endothermic reactions with O and OH.

In the present paper, we explore this problem one step further by
integrating {\it in parallel} for the detailed chemistry, ionization, and
thermal state along MHD disk wind streamlines. We use a model developed for
magnetized molecular interstellar shocks with an extended chemical network
of 134 species \citep{FlPi03b}, adding irradiation by (attenuated) coronal
X-ray photons and FUV photons from the accretion shock. Another difference
with respect to \citet{Safi93a} is the use of a slower and denser MHD
accretion-ejection solution that better matches the rotation signatures,
poloidal velocities, and ejection-accretion ratio in atomic T Tauri jets
\citep{Pes04,Ferr06}. The flow dynamics, thermo-chemistry, and treatment of
FUV and X-ray radiation, are described in detail in Section~\ref{s:model}, which may be skipped in a first reading. Our results are presented in Section~\ref{s:results} for a range of accretion rates and stellar masses representative of the Class~0, Class~I, and most active Class~II phases of solar-type stars. In Section~\ref{s:discussion}, the results are compared to static irradiated disk models and to observational trends, and model limitations are discussed. Conclusions are summarized in Section~\ref{s:concl}.

\section{Model description}\label{s:model}
\subsection{Dynamical MHD solution}\label{s:dynamics}
The adopted dynamical MHD model sets the total density, velocity, magnetic field and current in space. A large scale magnetic field of bipolar topology is assumed to thread the disk on a large radial extension. The wind model satisfies the full single-fluid MHD equations and has been obtained by assuming steady-state, axisymmetry, and a self-similar functional form for all physical variables. These models are an extension of the \citet{BlPa82} ideal MHD jet models in that they also solve for the full 3D dynamics of the underlying resistive keplerian accretion disk (where accreting matter needs to cross field lines), including {\it all} dynamical terms in the vertical, azimuthal, and radial directions. In particular, the disk compression and (slight) sub-keplerian rotation induced by the magnetic field, addressed in \citet{shu08}, are accurately treated. The accretion-ejection transition, wind mass-loading, and slow point crossing, are thus all determined self-consistently, given a vertical profile of temperature and resistivity in the disk (see \citealt{Ferr97, CaFe00} for details). 

A key free parameter of the models that is most amenable to observational constraints is $\xi$, the disk {\it ejection efficiency}, fixing the accretion rate as a function of the radius as $\Macc\propto r^\xi$. The total mass outflow to disk accretion rate over a radial extension $\rin$ to $\rout$ is then $2\dot M_{\rm j}/\Macc=\xi\ln(\rout/\rin)$. The parameter $\xi$ is also linked to the jet physics. As long as wind magnetic torques dominate, it is inversely related to the \citeauthor{BlPa82} magnetic lever arm parameter $\lambda$ through:
  \begin{equation}
  \lambda \equiv \frac{\rA^2}{r_0^2} \simeq 1 + \frac{1}{2\xi},
  \label{eq:lambda-xi} \end{equation}
where $r_0$ is the midplane radius of the magnetic surface along which the matter flows and $\rA$ is its cylindrical radius at the poloidal Alfv\'en point \citep{Ferr97}. The maximum asymptotic jet velocity and specific angular momentum along each streamline are given respectively by $\vp^\infty\simeq\sqrt{G\Mstar/r_0}\sqrt{2\lambda-3}$ and $r v_\phi^\infty\simeq\lambda\sqrt{G\Mstar r_0}$ \citep{BlPa82}, and are thus smaller for decreasing $\lambda$ and increasing ejection efficiency $\xi$.

In \citet{Garcia01a}, we calculated the thermal-ionisation structure for {\it atomic} MHD disk wind solutions with high lever arms ($42\lesssim\lambda\lesssim84$), obtained for a vertically isothermal disk profile \citep{Ferr97}. \citet{Garcia01b} presented synthetic forbidden line predictions and showed that the predicted line centroid velocities were on average too high when compared to observations, and that models with lower lever arms should be favored. This conclusion was later confirmed by the small rotation signatures observed in T Tauri atomic jets, which rule out values of $\lambda\geq20$ \citep{Pes04,Ferr06}.
 
\begin{figure}\centering
  \includegraphics[width=0.9\columnwidth]{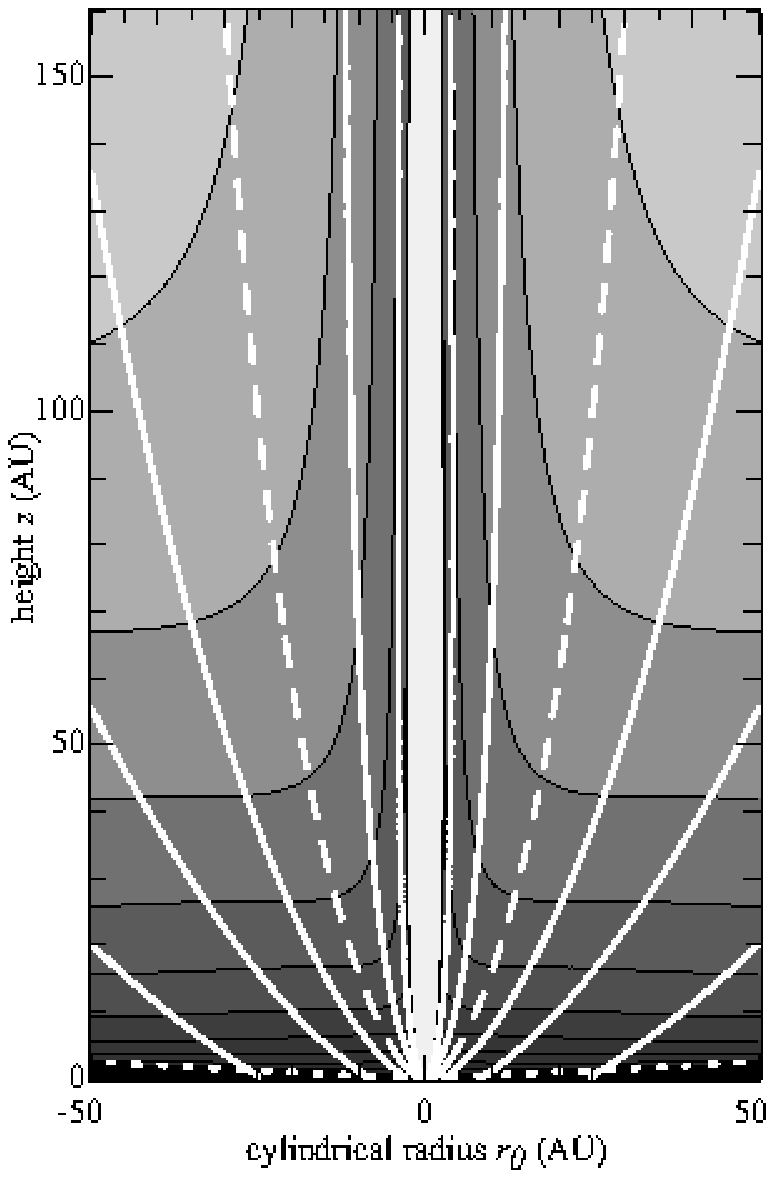}
  \caption{Overall geometry of the "slow" MHD disk wind solution of \citep{CaFe00} used in this article. Solid white curves show various magnetic flow surfaces, with that anchored at 1~AU shown in dashed. The density for $\Macc=10^{-6}\Msun/{\rm yr}$ and $\Mstar=0.5\Msun$, is coded in the contour plot starting at $8\times10^4~\cm^{-3}$ and increasing by factors of 2. The bottom dotted line traces the slow magnetosonic surface (at 1.7 disk scale heights) where our chemical integration starts. A color version of this figure is available in the online edition of this journal.}
  \label{f:model} \end{figure}

Therefore, in the present study we use a "slow" disk wind solution with a smaller lever arm parameter $\lambda=13.8$, developed by \citet{CaFe00}. This particular value was chosen as best reproducing the possible rotation signatures reported across the outer layers of the DG Tau atomic jet, with an inferred outer launch radius $\rout\simeq3$~AU (see \citealt{Pes04}). The same model can also account for the highest velocity $\simeq350~\kms$ detected on the jet axis, if launching occurs down to the corotation radius $\rin\simeq0.07$~AU \citep{Amb11}. Indeed, the predicted asymptotic speed in this model varies with anchor radius $r_0$ as
  \begin{equation}
  \vp^\infty\simeq100\left(\frac{r_0}{1\,{\rm AU}}\right)^{-1/2}
  \left(\frac{\Mstar}{0.5\Msun}\right)^{1/2}~\kms.
  \label{eq:vp-r0} \end{equation}
The corresponding ejection efficiency, $\xi=0.04$, also appears compatible with the observed ejection to accretion ratio in the DG Tau atomic jet, given current uncertainties \citep{Amb11}. The resulting ($r,z$) distribution of density and streamlines is shown in Fig.~\ref{f:model}. Thedensity contours turn from horizontal to vertical at polar angles \mbox{$\theta\leq18^\circ$}, producing an apparent density collimation in good agreement with observed jet widths \citep{Cab99,Garcia01b,RayPPV}. This dense axial beam is surrounded by wider streamlines from larger $r_0$ (white curves in Fig.~\ref{f:model}) that recollimate on larger scale and reach a lower terminal speed (see eq.~\ref{eq:vp-r0}). Such a steep transverse velocity decrease from axis to edge is observed across several resolved Class~II jets including DG~Tau and CW Tau \citep{lav00,bac00,cof07} and seems in good agreement with the chosen MHD disk wind solution \citep{Pes04,Cab09}.

"Slow" MHD solutions ($\lambda<20$) that become super Alfv\'enic may be obtained by including a small entropy deposition at the disk upper layers $z\simeq h$ \citep{CaFe00}. The warm mass-loading region is denser and the ejection efficiency $\xi$ is substantially enhanced, lowering the $\lambda$ value (see eq.~\ref{eq:lambda-xi}). Such additional heating was assumed to be powered by dissipation of MHD disk turbulence and parametrized as a tiny fraction $f$ of the disk accretion power, with $f=8\times10^{-4}$ in the solution chosen here (see \citealt{CaFe00} for a discussion of the effect of $f$ on the wind dynamics). This new class of solutions, referred to as "warm" in \citet{Pes04}, is still "cold" in a dynamical sense as the initial thermal energy remains negligible compared to gravity, and the wind acceleration is still mostly magnetic. Therefore, the wind temperature beyond the slow point does not affect the dynamics and we may recompute it {\it a posteriori} from the actual heating/cooling terms, without loss of self-consistency (see discussions in Section~\ref{s:discussion} and Appendix~\ref{app:heating}).

Other input properties of the adopted solution are: a thermal (uncompressed) aspect ratio $\varepsilon=h/r=C_s/V_K=0.03$ (with $C_s$ the sound speed and $V_K$ the kepler speed), a turbulent resistivity in the disk midplane $\nu_{\rm m}=\alpha_{\rm m}V_{\rm A,0}h$, with $V_{\rm A,0}$ the Alfv\'en velocity and $\alpha_{\rm m}=2$ (in order to ensure stationarity), and a turbulent viscosity $\nu_v=\nu_m$. Crossing of the slow magnetosonic point is obtained for a ratio of magnetic to thermal pressure in the disk midplane $\mu\simeq0.33$ (leading to quasi-keplerian rotation; \citealt{shu08}). The computed second \citeauthor{BlPa82} jet parameter is $\kappa\simeq0.1$, and the field inclination at the disc surface is $\theta_0\simeq57.3^\circ$, with similar surface values of $B_z$, $B_\phi$ and $B_r$.

For a given self-similar MHD solution, all physical quantities at a given polar angle obey specific scalings with the stellar mass $\Mstar$, the disk accretion rate $\Macc$, and the anchoring radius $r_0$ of the magnetic surface in the midplane, as given in eq.~(9) of \cite{Garcia01a}. In particular, the midplane field in the adopted MHD solution is,
  \begin{equation}
  B_z\simeq0.7G\left(\frac{r_0}{1\,{\rm AU}}\right)^{-5/4}
  \left(\frac{\Macc}{10^{-6}\Msun {\rm yr}^{-1}}\right)^{1/2}
  \left(\frac{\Mstar}{0.5 \Msun}\right)^{1/4},
  \end{equation}
which follows the same scaling as the available magnetic field measurements in protostellar disks \citep{shu-IAU243}. The required field strengths thus appear plausible. In the following, we will vary $\Mstar$, $\Macc$, and $r_0$ in order to explore their effect on the molecular content and thermal state of the disk wind.

\subsection{Thermo-chemical evolution}
The thermo-chemical evolution along wind streamlines was implemented by adapting the last version \citep{FlPi03b} of a code constructed to calculate the steady-state structure of planar molecular MHD multi-fluid shocks in interstellar clouds \citep{FlPH85}. The ion-neutral drift, FUV field, and dust-attenuation were calculated following the methods developed by \citet{Garcia01a} in the atomic disk wind case. Irradiation by stellar X-rays was also added, following the approach of \cite{SGSL02}.

Given the high densities and low ion-neutral drift speeds in the wind (see Section~\ref{s:results}) the same temperature and velocity is adopted here for all particles. The latter is prescribed by the underlying single-fluid MHD wind solution. We thus integrate numerically the following differential equations on mass density $\rho$, species number density $n(a)$, and temperature $T$ along the streamline, as a function of altitude $z$ above the disk midplane:
  \eqab \frac{\dr\rho_f}{\dr z} &=&
     \frac{S_f-\rho_f\nabla\cdot\vb}{v_{z}}, \\ 
     \frac{\dr n(a)}{\dr z} &=& \frac{R_a - n(a)\nabla\cdot\vb}{v_{z}},
     \label{eq:n_sp} \\
     \frac{\dr T}{\dr z} &=& \frac{\Gamma-\Lambda-n_{\rm tot}
     k_BT\nabla\cdot\vb -\frac{3}{2}k_BT R}{\frac{3}{2}k_Bn_{\rm
     tot}v_{z}}. \eqae
Here, $v_z$ and ($\nabla\cdot\vb$) are the bulk vertical flow speed and the (3D) flow divergence interpolated from the MHD wind solution, $n_{\rm tot}$ is the total number density of particles, $T$ is the temperature, $k_B$ is the Boltzmann constant, $S$ and $R$ are the rates of change in mass and number of particles, respectively, per unit volume, and $\Gamma$ and $\Lambda$ are the heating and cooling rates per unit volume. The equations on $n(a)$ apply to each species $a$ as well as to the individual populations of the first 49 levels of H$_2$ (up to an energy of $20\,000$~K) which are integrated in parallel with the other variables. The equations on $\rho_f$ apply to each "fluid" $f$ (neutral, positive, negative) and are used mainly for internal checking purposes, as the total mass density $\rho(z)$ is prescribed by the MHD solution.

Cooling and heating mechanisms include 
  \begin{itemize}
  \item Radiative cooling by H$_2$ lines excited by collisions with H,
  H$_2$, He, and electrons \citep{lebourlot99}.
  \item Radiative cooling by CO, H$_2$O, and $^{13}$CO in the Large Velocity
  Gradient approximation \citep{NeKa93}, and by OH and NH$_3$ in the
  low-density limit \citep{FlPH85}.
  \item Atomic cooling by fine-structure and metastable lines of C, N, O, S,
  Si, C$^+$, N$^+$, O$^+$, S$^+$, Si$^+$ \citep{Fletal03} and Fe$^+$
  \citep{Gian04}.
  \item Inelastic scattering of electrons on H and H$_2$
  \citep{ABBK91,Humm63,RaEn65}.
  \item Energy released by collisional ionisation and dissociation and
  exo/endo-thermicity of chemical reactions \citep{FlPH85}.
  \item Energy heat/loss through thermalization with grains
  \citep{Tielens85}.
  \item Ambipolar-diffusion heating by elastic scattering between the
  neutral fluid and charged ions and grains (\citealt{Garcia01a}, see
  Section~\ref{s:ambip}).
  \item Ohmic heating arising from the drift between electrons and other
  fluids (\citealt{Garcia01a}, see Section~\ref{s:ambip}).
  \item Photoelectric effect on grains \citep[][eq.~42]{BaTi94} irradiated
  by the (attenuated) FUV field of hot accretion spots (see
  Section~\ref{s:UV}).
  \item Heating through cosmic-rays and (attenuated) coronal stellar X-rays
  (\citealt{DaYL99}; see Section~\ref{s:xrays}).
  \end{itemize}

The reader is referred to the corresponding references for a discussion of the physical context and hypotheses involved in modelling each of the above processes.

\subsubsection{Chemical network}
The chemical network consists of 134 species, including atoms and molecules (either neutral or singly ionized) as well as their correspondents inside grain refractory cores, and on grain icy mantles. A representative polycyclic aromatic hydrocarbon (PAH) with 54 carbon atoms is also included, with a fractional abundance of $10^{-6}$ per H nucleus. The total elemental abundances and their initial distribution among gas, grain cores, and icy mantles are taken from Tables~1 and 4 of \citet{FlPi03b}.

The charge balance of grains and PAHs is treated as in \cite{FlPi03b}. The number of grains per H nucleus and the mean grain cross section are determined from the abundance of depleted elements in grain cores, and the adopted grain size distribution (see Section~\ref{s:Av}). Most of the grains are charged, and assumed well-coupled to the charged fluid.

We consider 1143 reactions including neutral-neutral and ion-neutral reactions, recombination with electrons, charge exchange, cosmic ray induced desorption from grains, sputtering of grain icy mantles, and erosion of charged grain cores by impact of drifting heavy neutral species. Reaction rates between charged and neutral species are enhanced by ion-neutral drift, following the effective temperature prescription of \cite{FlPH85}, with the drift speed $\vb_{in}$ computed as described in Section~\ref{s:ambip}.

H$_2$ collisional dissociation is treated level by level following \citet{lebourlot02}. Reformation on grains is computed with the sticking coefficient of \citet{sticking}. Three-body reactions in the gas phase are not included, therefore we will limit ourselves to densities $\le10^{12}$ cm$^{-3}$.

Ionisation and dissociation reactions by far-ultraviolet (FUV) photons from
accretion hot spots, and by stellar coronal X-rays, are included as
described in Sections~\ref{s:UV} and \ref{s:xrays}, respectively.

\subsubsection{Initial conditions and integration}\label{s:initial}
The integration of the set of equations for temperature, mass and number of particles along the flow is an initial value problem. Thus initialization of temperature and initial populations have to be devised. As in \citet{Garcia01a}, we start all calculations from the wind slow magnetosonic point (located at $1.66h(r)$ for our adopted MHD solution), and assume that the temperature and species abundances there are at equilibrium with the local radiation field. These equilibrium values are obtained by performing a "steady-state" run over $10^5$ yrs where thermo-chemical equations are solved with the density held fixed. We then integrate the thermo-chemical equations along the flow using the DVODE integrator \citep{dvode}, until the recollimation point where the streamline reaches its maximum radial extension (radius of $40r_0$ at $z=900r_0$ for our chosen wind solution). We checked that the final temperature and H$_2$ abundance along the streamline do not depend sensitively on the initial equilibrium conditions, as long as the gas is fully molecular initially. This occurs because the ionisation (which controls the ion-neutral drift heating) adjusts rapidly at the dense wind base to the local radiation field, even though it becomes frozen-in at large distances.

\subsection{Ambipolar diffusion coefficients}\label{s:ambip}
The ambipolar diffusion and Ohmic heating terms require a detailed calculation of elastic momentum exchange rates and drift speeds. The ion-neutral drift speed is calculated from the MHD solution using the analytical formula derived in Appendix~A of \cite{Garcia01a}, valid for $||\vb_{in}||\ll||\vb||$:
  \eqab \vb_{in} &=& \frac{1}{1+p}\left(\frac{\Jb\times\Bb}
  {c(1+X)\overline{\mu_{in}n_i\nu_{in}}}+p\frac{\Jb}{q_en_e}\right)
         \label{eq:velin} \\
  p &=& \frac{\overline{\mu_{en}n_e\nu_{en}}}
             {\overline{\mu_{in}n_i\nu_{in}}} \eqae
where $\Jb$ is the electric current density, $\Bb$ is the magnetic field, $c$ is the speed of light, $q_e$ is the electron charge, $n_e$ and $n_i$ are the number densities of electrons and ions, $X=\rho_i/\rho_n$, and $\overline{\mu_{in}n_i\nu_{in}}$  and $\overline{\mu_{en}n_e\nu_{en}}$ are the total momentum transfer rate coefficients for ion-neutral and electron-neutral collisions, respectively.

The total ion-neutral momentum transfer rate coefficient is obtained by summing over the main neutrals (H, H$_2$, He) and over all ions, charged PAHs and charged grains:
  \eqb \overline{\mu_{in}n_i\nu_{in}} =
       \sum_{\begin{array}{c}a=\mathrm{ions},\\
       \mathrm{PAH}^{\pm},g^{\pm}\end{array}}
       \sum_{\begin{array}{c}b=\\\Hr,\Hr_2,\He\end{array}}
       \mu_{ab}n(a) n(b) \left<\sigma v\right>_{ab}, \label{e:muinninuin}
       \eqe
where $n(a)$ is the number density of species $a$, $\mu_{ab}$ is the reduced mass of particles $a$ and $b$, and $\left<\sigma v\right>_{ab}$ is their elastic collision rate coefficient; the latter is evaluated via the recent analytical fits to quantum-mechanical calculations provided in Table~2 of \citet{PiGa08}, when available, given as a function of the effective relative speed $\bar{v}$:
  \eqab
  \bar{v} &=& \sqrt{\frac{8k_B T_r}{\pi\mu_{ab}}+ ||\vb_{in}||^2}
  \label{e:vbar}\\ {\rm with}\quad
        T_r &=& \frac{m_aT_n+m_bT_i}{m_a+m_b}. \eqae 
For the H$_2$-H$^+$ pair we use the updated fit in \cite{PiGa08err}. For the rest of the pairs $\left<\sigma v\right>_{ab}$ is taken as the maximum of the polarizability and hard sphere rate coefficients \citep{Garcia01a}:
  \eqb \left<\sigma v\right>_{ab} =
  \max\left\{2.41\pi q_e\sqrt{\frac{\alpha_b}{\mu_{ab}}},
             \sigma_s\bar{v}\right\}, \eqe
where $\alpha_b$ is the polarizability of H, He, or H$_2$ and the hard sphere cross section is taken as
  \eqb \sigma_s = \left\{
  \begin{array}{lll}10^{-14}~\cm^2&,&\mathrm{PAH}^{\pm}\\
  \pi\left<R_d^2\right>&,&\mathrm{grains}\ (g^\pm) \\
  10^{-15}~\cm^2&,&\mathrm{else}\end{array}\right. \eqe 
Here $\pi\left<R_d^2\right>$ denotes the average grain cross section over our size distribution. The total momentum transfer rate coefficient for collisions between electrons and neutrals is given by:
  \eqb \overline{\mu_{en}n_e\nu_{en}} =
  m_en_e\sum_{a=\Hr,\Hr_2,\He}n(a) \left<\sigma v\right>_{ea} \eqe
with the $\left<\sigma v\right>_{ea}$ also evaluated from the recent analytical fits of \cite{PiGa08}. The ion-electron momentum transfer rate entering Ohmic heating is calculated with the classical formula of \citet{Schunk75} (see eq.~A.13 of \citealt{Garcia01a}).

\begin{figure}\centering
  \includegraphics[scale=.73]{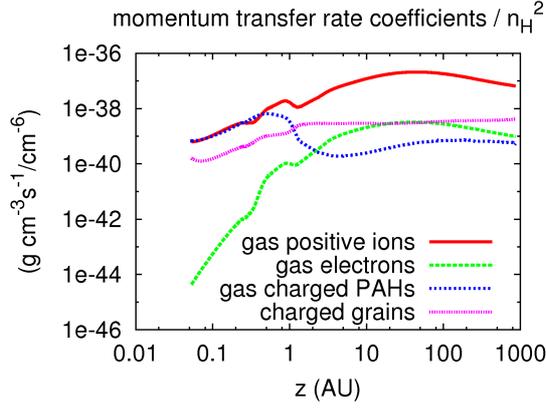}
  \caption{\label{f:MTRCs}Momentum transfer rate coefficients (normalized by
  $n_\Hr^2$) in collisions of the neutral fluid with the different charged
  fluids as a function of altitude along a flow line launched at 1~AU, for
  \mbox{$\Macc=10^{-6}\Msun$ yr$^{-1}$}, $\Mstar=0.5\Msun$ (our Class~I
  model). A color version of this figure is available in the online edition
  of this journal.}
  \end{figure}

Fig.~\ref{f:MTRCs} plots the momentum transfer rate coefficients (normalized by $n_\Hr^2$) for collisions between the neutral fluid and the different charged fluids as a function of vertical distance above the disk midplane for a representative streamline. It may be seen that collisions with positive ions dominate the coupling, with an extra contribution from charged PAHs near the flow base.

The ambipolar drag heating and Ohmic heating are given by \citep{Garcia01a}:
  \eqb
  \Gamma_\mathrm{drag} = \left(\frac{{\rho_n}}{\rho}\right)^2
  \frac{\frac{1}{c^2}\Vert\vec{J}\times\vec{B}\Vert^2}
            {\overline{\mu_{in}n_i\nu_{in}}}, \quad
  \Gamma_\mathrm{ohm}={\eta}\Vert\vec{J}\Vert^2 ,
  \label{eq:gdrag}\eqe
with $\eta=(\overline{\mu_{ie}n_i\nu_{ie}}+\overline{\mu_{en}n_n\nu_{ne}})/(q_en_e)^2$. Note that $\Gamma_\mathrm{drag}$ is proportional to the square of the Lorentz force per particle, and inversely proportional to the ionisation fraction in the wind.

\subsection{FUV field}\label{s:UV}
The FUV radiation will control the ion abundance (and therefore the ion-neutral drift) in the upper wind regions. It will also photodissociate molecules, and provide extra heating by collisions with warm dust and photoelectric effect on grains. Our treatment essentially follows the approach of \citet{Garcia01a} with several updates, including an accretion spot geometry and a photochemical network.

\subsubsection{Unattenuated radiation field}\label{s:UVfield}
Instead of the optically thick equatorial boundary layer model adopted in \cite{Safi93a} and \cite{Garcia01a}, we adopt here the paradigm of magnetospheric accretion, recently favored in T Tauri stars, where the inner disk is truncated at several stellar radii and accreting material flows along field lines in quasi free-fall, giving rise to shocked "hot spots" on the stellar surface \citep{bouvier}. The accretion geometry in younger protostars is not yet constrained observationally, but their similar X-ray flare properties to T Tauri stars indicate an already well-developed stellar magnetosphere \citep{Imanishi03}, hence we will adopt the same paradigm for simplicity.

The FUV radiation field thus comes from the central low-mass star of effective temperature $\Tstar=4\,000$~K, and from hot accretion spots of fixed temperature $T_\hs=10\,000$~K, chosen to match the mean observed colour temperature of the FUV excess in T Tauri stars \citep{uv-slope}. Both radiation sources are treated as black bodies.

As in \citet{Garcia01a}, we neglect the scattering contribution to the FUV field%
  \footnote{Radiative transfer calculations in 1+1D show that it starts
  being important only very near the disk plane \citep{nomura}. Ionization
  and dissociation in this region will be dominated by hard stellar X-rays
  in our model.}.
The (unattenuated) direct stellar flux at distance $R$ from the star (with dimensions erg cm$^{-2}$ s$^{-1}$ Hz$^{-1}$) is calculated through the relation
  \eqb F_\nu(R) = \left(B_\nu(\Tstar)+B_\nu(T_\hs)
  \delta_\hs \right) \frac{\pi\Rstar^2}{R^2} \label{e:flux}\eqe
where $B_\nu(T)$ denotes the specific intensity at frequency $\nu$ of a black body of temperature $T$, $\Rstar$ is the stellar radius, and $\delta_\hs$ is the fraction of the total stellar surface $4\pi\Rstar^2$ covered by hot spots (assumed uniformly distributed). The main difference with the boundary-layer model of \citet{Garcia01a} is thus a fixed $T_\hs$ independent of $\Macc$, and a more isotropic radiation flux (due to the lack of projection effects or disk occultation).

In each model, $\delta_\hs$ is determined by requiring that the hot spots
radiate half of the accretion luminosity,
i.e.~$4\pi\Rstar^2\delta_\hs\sigma_BT_\hs^4=G\Macc\Mstar/(2\Rstar)$. The
value of one half is meant to be illustrative only: the actual fraction of
accretion luminosity radiated in the accretion shock could exceed 80\% for a
large disk truncation radius $>5\Rstar$, or decrease if a sizeable fraction
of accretion energy is tapped to drive a stellar wind. None of these effects
being well quantified, especially in Class 0 and Class I sources, we adopt
50\% here for illustration and consistency with the boundary-layer model of
\citet{Garcia01a}. We set $\Rstar=3\Rsun$, the radius of a young accreting
$0.5\Msun$ star near the "birthline" with $\Tstar=4\,000$~K according to the
models of \citet{Stahler88}, noting that it remains a good approximation for
a very young Class 0 ($\Rstar\simeq 2\Rsun$ for a $0.1\Msun$ protostar
accreting at $5\times10^{-6}\Msun\yr$; cf.~\citealt{Stahler88}). With the
above assumptions, the modelled FUV hot spot continuum in the domain
$1\,400-2\,000$~{\AA} is a factor $1.5-2.5$ stronger than the accretion
shock models of \citet{Gullbring00} for DR~Tau and DG~Tau, using the same
accretion rates ($\Macc=3-5\times10^{-7}\Msun$ yr$^{-1}$). On the other
hand, if the FUV spectrum below $1\,400$~{\AA} were as flat as in the lower
accretion stars BP~Tau and TW Hya \citep{Bergin03}, our blackbody model
would underestimate the flux at $1\,000$~{\AA} by a factor $4-2$. This
remains sufficiently accurate for the present exploratory study.

Since we do not solve for radiative transfer, we neglect the extra FUV flux in the stellar Ly$\alpha$ line. Ly$\alpha$ pumping of highly excited levels of H$_2$ is occuring close to the star \citep{Herczeg02,herczeg06,nomura} but will be less important on the larger jet scales of interest here, especially since we focus on dense molecular jets where H$_2$ is well shielded.

\subsubsection{Attenuation by dust}\label{s:Av}
\begin{figure}\centering
  \includegraphics[scale=.73]{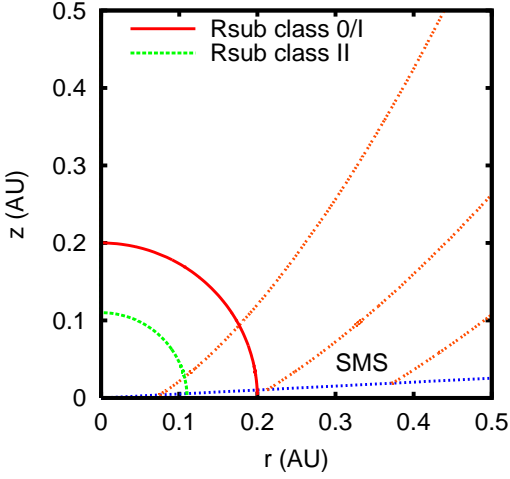}
  \caption{The dust sublimation surface in our Class~0/I models (outer red circle) and our Class~II model (inner green circle), for the grain size distribution and mixture of silicates and graphite proposed by \cite{MaRN77}. The wind slow magnetosonic surface (denoted SMS) is superimposed as well as three flow lines (orange dotted lines), including the innermost one launched from 0.07~AU. The stellar parameters are $\Tstar=4\,000$~K and $\Rstar=3\Rsun$, while the hot spot temperature and coverage fraction are $T_\hs=10\,000$~K, and $\delta_\hs=0.032$ (Class~0/I) and 0.0032 (Class~II). A color version of this figure is available in the on-line edition of this journal.}
  \label{f:Rsubs} \end{figure}

The stellar FUV field is attenuated mainly by dust in the disk wind along the line of sight to the star. The sublimation radius $\Rsub$ is calculated via eq.~(B.5) of \cite{Garcia01a}, considering a unique sublimation temperature $T_{\rm sub}=1\,500$~K. We used the dust optical properties tabulated by \cite{DrLe84,DrMa93,LaDr93}, assuming the standard "MRN" mixture of astronomical silicates and graphite and the grain radius distribution $n(a) \propto a^{-3.5}$ proposed by \cite{MaRN77}, with grain radii in the interval [0.005,0.25] $\mu$m. The sublimation surface for the Class~0/I/II models in Section~\ref{s:results} is plotted in Fig.~\ref{f:Rsubs} on top of inner disk wind streamlines.

The dust optical depth to the star at point of spherical coordinates $(R,\theta)$ is $\tau_\nu(R,\theta)=\overline{\sigma_d(\nu)}N_\Hr^d(R,\theta)$ where $\overline{\sigma_d(\nu)}$ is the dust extinction cross section per H nucleus (integrated over the grain size distribution), and $N_\Hr^d(R,\theta)$ is the column density of hydrogen nuclei on {\it dusty} wind streamlines along the line of sight to the star. In our self-similar wind model, the wind density varies as $R^{-1.5}f(\theta)$, and $N_\Hr^d(R,\theta)$ may then be integrated analytically as:
  \eqb
  N_\Hr^d(R,\theta)=2n_\Hr(R,\theta)R
  \left(\sqrt{\frac{R}{\Rmin(\theta)}} - 1\right). \label{eq:nh} \eqe
Here, $n_\Hr$ is the number density of H nuclei in all forms, \mbox{$n_\Hr\simeq n(\Hr)+2n(\Hr_2)+n(\Hr^+)$}, and $\Rmin(\theta)$ is the spherical radius inside which there is no dust, i.e.~either the sublimation radius, $\Rsub$, or the radius of the innermost flow line at angle $\theta$, $\Rin(\theta)$, whichever is larger (see Fig.~\ref{f:Rsubs}). The innermost wind streamline is assumed launched from $\rin=0.07$~AU, a typical corotation radius in young low-mass stars.

The total dust geometrical cross section per H nucleus, calculated from our grain size distribution and the abundance of depleted elements in grain cores, is $\pi\left<R_d^2\right>n_d/n_\Hr=1.36\times10^{-21}~\cm^2$, where $\left<R_d^2\right>$ is the mean square grain radius; the ratio of $A_V$ to $N_\Hr$ is then $\overline{\sigma_d(5500~\AA)}=2.72\times10^{-22}~\cm^2$.

\subsubsection{Dust temperature}\label{s:Tdust}
Collisions with dust grains heated by the UV field can be an important gas heating/cooling term at the wind base (though unimportant further out). We assume that above the wind slow point ($z\geq1.7h$), dust is optically thin to its own radiation. The dust temperature is then calculated at every step from the grain radiation equilibrium against incident (wind-attenuated) stellar photons:
  \eqb 4\pi\sigma_BT_d^4 \left<R_d^2\right>
  \cdot\overline{\left<Q_\emr(T_d)\right>} =
  \int_0^\infty\overline{\sigma_d(\nu)}F_\nu(R)
  \er^{-\overline{\sigma_d(\nu)}N_\Hr^d(R,\theta)}\dr\nu \eqe
where $\sigma_B$ is the Stefan-Boltzmann constant, $\left<R_d^2\right>$ is the mean square grain radius, $\overline{\left<Q_\emr(T_d)\right>}$ is the Planck-averaged grain emission efficiency (weighted by $n(a)a^2$), and $\overline{\sigma_d(\nu)}$ is the dust absorption cross section per H nucleus defined earlier. Collisions with the gas are neglected in the grain thermal balance, as they compete with radiation only at densities $>10^{13}~\cm^{-3}$ \citep{Glass04}. We verified that when the FUV excess and the wind attenuation are negligible, our $T_d$ values agree with detailed calculations at the top of the disk atmosphere by \cite{dalessio99} for MRN dust and a $4\,000$~K star (taking into account differences in $\Rstar$).

\subsubsection{Photochemical network}\label{s:photochem}
A network of photoionisation and photodissociation reactions is implemented
with rates in the form $k_p=\chi\,\gamma\mathsf{e}^{-\beta A_V}$, where
$A_V$ is the visual extinction calculated above (Section~\ref{s:Av}),
$\beta,\gamma$ are constants depending on the reaction, taken from
\cite{vandishoeck} and \cite{roberge}, and $\chi$ is the ratio at
$1\,000$~{\AA} of the unattenuated FUV flux to the mean interstellar
radiation field of \citet{Drai78}, integrated over
$4\pi(\mathrm{2\times10^{-6}\,erg~s^{-1}~cm^{-2}}$~\AA$^{-1})$. Such a
"monochromatic" flux normalization at $1\,000$~{\AA} was chosen rather than
the average ratio to the interstellar field over the interval
$910-2\,066$~\AA, denoted $G_0$ in the literature (see
e.g.~\citealt{Tielens85}), as photoreactions contributing to the main
ionisation and dissociation reactions in our models (C, S, CH$^{+}$, H$_2$,
CO) occur in the narrow range $910-1\,200$~\AA.

Special cases of photodissociation reactions are the dissociation of H$_2$ and CO, which occur through line absorption. We thus need to consider not only the shielding caused by the dust particles, but also the additional self and cross shielding due to those species themselves. They were implemented in the model in an approximate way as described below.

\subsubsection{H$_2$ photodissociation.}\label{s:H2photo}
We adopted an H$_2$ photodissociation rate coefficient per molecule of the form proposed by \cite{DrBe96}:
  \eqb k_{p,\Hr_2}=\chi\,p_0({\Hr_2})\,
  \times f_{\Hr_2,d}(A_V)f_{\Hr_2,\Hr_2}(N_{\Hr_2},b)\eqe
where $p_0({\Hr_2})=4.2\times10^{-11}$~s$^{-1}$ is the unshielded rate in
the Draine field, $f_{\Hr_2,d}(A_V)=\er^{-\beta_d A_V}$ is the shielding
factor by dust, with
\mbox{$\beta_d=\sigma_d({1000~\AA})/\sigma_d({5500~\AA})=6.3$} for our dust
mixture, and $f_{\Hr_2,\Hr_2}(N_{\Hr_2},b)$ is the H$_2$ self-shielding
factor given by eq.~(37) in \cite{DrBe96}, with the line width parameter $b$
set here to the local sound speed $C_s$. The shielding column of H$_2$
molecules on the line of sight to the star, $N_{\Hr_2}$, depends on the
H$_2$ abundance on all inner wind streamlines. In order to allow a local
calculation in the present exploratory study, we assume a mean H$_2$
abundance on the line of sight to the star equal to half the local H$_2$
abundance at the current point of the streamline. This "local" assumption
tends to overestimate self-shielding, but will not affect our results when
photodissociation is not the dominant destruction mechanism. A conservative
check of this hypothesis is made a posteriori (see
Section~\ref{s:results}).

\subsubsection{CO photodissociation.}\label{s:COphoto}
\begin{figure*}[t] \centering
  \includegraphics[width=\textwidth]{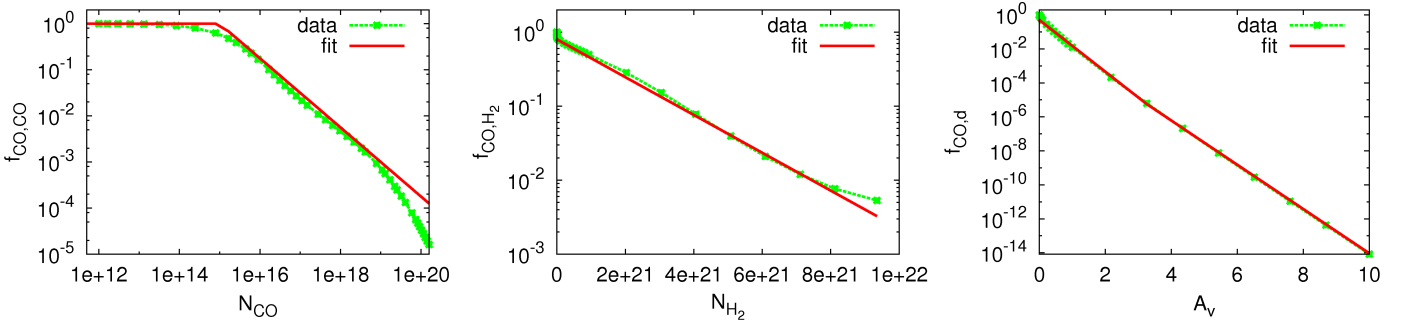}
  \caption{Adopted analytical fits (red curves, from eq.~\ref{eq:co-shield}) to the CO photodissociation shielding factors of \cite{LHPR96} (connected green symbols) caused by CO line overlap (left), H$_2$ line overlap (middle) and dust extinction (right).}
  \label{f:co-shielding} \end{figure*}

For CO, one needs to consider not only the self-shielding and the shielding by dust, but also the shielding by H$_2$ (through line overlap with CO). We have fitted the data of Table~11 from \cite{LHPR96} as a function of $A_V$ and of the shielding columns $N_{\mathrm{CO}}$ and $N_{\mathrm{H_2}}$, and obtained the following analytical formulae for the shielding factors by dust, CO, and H$_2$ respectively:
  \eqab f_\mathrm{CO,d}(A_V) &=& \max\left\{0.5\er^{-3.5A_V},
  0.1\er^{-3A_V}\right\}, \\ f_\mathrm{CO,CO}(N_{\mathrm{CO}}) &=&
  \min\left\{1,
  \left(\frac{N_\mathrm{CO}}{10^{15}~\cm^{-2}}\right)^{-0.75}\right\}, \\
  f_\mathrm{CO,H_2}(N_{\mathrm{H_2}}) &=&
  0.8\er^{-N_{\Hr_2}/1.7\times10^{21}~\cm^{-2}}.
  \label{eq:co-shield}\eqae
These analytical formulas are valid to within a factor of 2 up to $N_\mathrm{CO}\simeq2\times10^{19}~\cm^{-3}$, and over the whole range in $N_{\Hr_2}$ and $A_V$, as shown in Figure~\ref{f:co-shielding}.

The photodissociation rate coefficient per molecule is then
  \eqb k_{p,{\rm CO}}=\chi p_0({\rm CO}) \times f_{\mathrm{CO},d}(A_V)
  f_{\mathrm{CO,CO}}(N_{\mathrm{CO}}) f_{\mathrm{CO},\Hr_2}(N_{\Hr_2}),\eqe
where $p_0({\rm CO})=2\times10^{-10}$s$^{-1}$ is the unshielded rate for an isotropic Draine field (twice the value from \citet{LHPR96}, who assume a one-sided cloud illumination). As done for H$_2$, the shielding column of CO molecules on the line of sight to the star, $N_{\mathrm{CO}}$, is calculated assuming a mean CO abundance on inner disk streamlines of half the local value of the CO abundance at the current point.

\subsection{X-rays}\label{s:xrays}
Hard coronal X-rays will be the dominant ionization process at the base of the dusty disk wind, where the stellar FUV flux is strongly extinguished by dense inner streamlines. Energetic secondary electrons generated by X-ray ionisation will also dissociate and heat the gas. Our adopted X-ray treatment and chemical network are described in the following.

\subsubsection{X-ray spectrum and attenuation}
\label{sec:xray-tau}
\begin{figure}\centering
  \includegraphics[scale=.73]{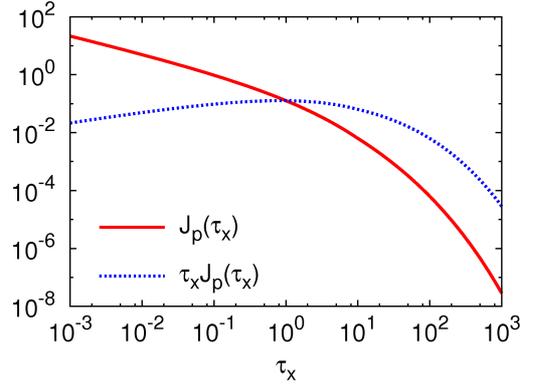}
  \caption{\label{f:Ip}The function $J_p(\tau_X)$ describing the attenuation of the energy deposition rate by a thermal X-ray spectrum versus $\tau_X$, the X-ray optical depth at the characteristic energy, $kT_X$. Also plotted is the function $\tau_X J_p(\tau_X)$ describing the effect of changing $kT_X$ (i.e.~$\tau_X$) for a fixed attenuating gas column density (see eq.~\ref{eq:ktx}).}
  \end{figure}

The X-ray flux is modelled as a thermal spectrum with characteristic energy $kT_X$,
  $$F_X(E) \equiv \dr L_X(E)/dE = (L_X/kT_X) \times \exp(-E/kT_X)$$
The attenuated rate of X-ray energy deposition {\it per H nucleus} in the wind writes 
  \eqb H_X (R,\theta)=\int^\infty_{E_0}\frac{F_X(E)}{4\pi R^2}\sigma_\pe(E)
  \exp^{-\sigma_\pe(E)N_\Hr(R,\theta)} \dr E,\label{e:hx} \eqe
where $E_0$ is the low energy cutoff of the incoming X-ray spectrum, $\sigma_\pe(E)$ is the photoelectric cross section per H nucleus, and $N_\Hr(R,\theta)$ is the total column of H nuclei to the star through inner wind streamlines (given by eq.~\ref{eq:nh} with \mbox{$\Rmin(\theta)=\Rin(\theta)$}). We calculate $H_X$ analytically following \cite{Glassgold97} and \cite{SGSL02}, who adopted a low-energy cutoff \mbox{$E_0=0.1$~keV} and a power-law approximation to the cross section, $\sigma_\pe(E)=2.27\times10^{-22}~\cm^{-2}~(\keV/E)^{p}$ with $p=2.485$. $H_X$ then writes
  \eqb {H_X}(R,\theta) = H_0(R) I_p(\xi_0,\tau_X)\label{e:hx0} \eqe
where
  \eqab H_0(R) &=& \frac{L_X}{4\pi R^2}{\sigma_\pe(kT_X)} , \\
  I_p(\xi_0,\tau_X) &=& \int_{\xi_0}^\infty\er^{-(\xi+\tau_X\xi^{-p})}
  \xi^{-p}\dr \xi \simeq J_p(\tau_X), \eqae
$\xi_0=E_0/kT_X$, $\tau_X\equiv\sigma_\pe(kT_X)N_\Hr(R,\theta)$ is the X-ray optical depth at $kT_X$, and $J_p(\tau_X)$ is an analytical fit given in eq.~(C3) of \citealt{SGSL02}, and plotted in Fig.~\ref{f:Ip}. The fit is valid for \mbox{$\tau_X>\tau_0=1/p\xi_0^p(p+\xi_0)=10^{-4}$}, a condition verified in our models.

The effect of changing $kT_X$ for a fixed $L_X$ and fixed position $(R,\theta)$ (and therefore a fixed attenuating column $N_\Hr$) may be visualised by rewriting $H_X$ as
  \eqb {H_X}(R,\theta) =  H_0(R)J_p(\tau_X) =
  \tau_X J_p(\tau_X) \times L_X/(4\pi R^2 N_\Hr). \label{eq:ktx} \eqe
The function $\tau_X J_p(\tau_X)$, also plotted in Fig.~\ref{f:Ip}, is almost flat over a broad range of $\tau_X$ (0.001 to 100). The regions where X-ray ionisation dominates in our models have values of $\tau_X\simeq0.3-30$ within this range for $kT_X=4$~keV. Furthermore, the ionization fraction varies only with the square root of ${H_X}$. Varying $kT_X$ in the range $2.5-15$~keV (i.e.~$\tau_X$ in the range $0.01-100$) would thus not strongly modify our results.

\subsubsection{X-ray ionization rates}
X-ray ionization rates are calculated with the same simplifying assumptions as in \cite{Glassgold97} and \cite{SGSL02}: The mean energy to form an $A^+$ ion is approximated by its constant high-energy limit $W(A^+)$, valid for primary photoelectrons with $E>E_0=0.1$~keV. Neglecting the small difference in energy between the absorbed X-ray photon and the primary photoelectron, and recalling that the absorbed energy $H_X$ is {\it per H nucleus}, the X-ray ionization rate of species $A$ {\it per atom} is then given by
  \eqb \zeta(A^+)= \frac{H_Xn_\Hr}{W(A^+)}\frac{1}{n(A)}\label{e:zetax}\eqe

The limiting mean energies $W(A^+)$ for X-ray ionization of H$_2$, H, and He
in neutral gas mixtures with 10\% of Helium have been calculated by
\cite{DaYL99} who fitted them%
  \footnote{The variation of $W(A^+)$ with ionization fraction is neglected
  here, as it amounts to less than 3\% for the ionization fractions
  $\le10^{-3}$ encountered in our models \citep{DaYL99}}
as linear functions of $n(\Hr)/n(\Hr_2)$ (middle terms of
eq.~\ref{wh2}-\ref{whe} below). We further simplify their expressions by
taking $0.53\simeq0.5$, $1.89\simeq2$, and recalling that the number density
of H nuclei $n_{\rm H}\simeq n(\Hr)+2n(\Hr_2)$ at low ionization, and
$n(\He)=0.1n_H$, which yields the right-hand side terms:
  \eqab
  W(\Hr_2^+) &=& 41.9 {\rm eV} \left[1 + 0.53\frac{n(\Hr)}{n(\Hr_2)}\right]
  \simeq 20 {\rm eV} \frac{n_\Hr}{n(\Hr_2)}, \label{wh2}\\
  W(\Hr^+) &=& 39.5 {\rm eV} \left[1 + 1.89\frac{n(\Hr_2)}{n(\Hr)} \right] 
  \simeq 40 {\rm eV} \frac{n_\Hr}{n(\Hr)}, \label{wh}\\
  W(\He^+) &=& 470 {\rm eV} \simeq 47.0 {\rm eV}\frac{n_\Hr}{n(\He)}.
  \label{whe}
  \eqae

Inserting the right-hand side expressions for $W(A^+)$ in eq.~\eqref{e:zetax}, the X-ray ionization rates {\it per specie $A$} become independent of $n_{\rm H}$ and $n(A)$ and may be written solely in terms of the parameter $\zeta(\Hr_2^+)$ that enters our chemical network: 
  \eqab \zeta(\Hr_2^+) &\equiv& H_X / 20 {\rm eV}, \label{e:zetah2}\\ 
  \zeta(\Hr^+) &\simeq& 0.5 \zeta(\Hr_2^+), \\ 
  \zeta(\He^+) &\simeq& 0.4 \zeta(\Hr_2^+).
  \eqae 
The total ionization rate {\it per H nucleus}, often denoted $\zeta_X$ in the literature, is related to $ \zeta(\Hr_2^+)$ by
  \eqab 
  \zeta_X &=&  \zeta(\Hr_2^+) \frac{0.5n(\Hr)+n(\Hr_2)+0.4n(\He)}{n_H}\\
          &=& 0.54 \zeta(\Hr_2^+),
  \eqae 
regardless of the ratio H/H$_2$. The formation of H$^+$ through dissociative ionization of H$_2$ is also included with a rate of $\zeta(\Hr_2^+)/22$ per molecule \citep{DaYL99}. A constant cosmic ray ionization rate $\zeta_{\rm CR}(\Hr_2^+)=5\times10^{-17}$ s$^{-1}$ is added to the rate produced by stellar X-rays, but it plays a negligible role in our models.

\subsubsection{X-ray induced dissociation}
The rate of H$_2$ dissociation by X-ray induced electrons is taken as $0.5\zeta(\Hr_2^+)$, following the results of \cite{DaYL99} for primary electron energies $>0.1$~keV. The energetic electrons also collisionally excite H$_2$ to higher electronic levels, which then radiatively decay by emitting a flux of ultraviolet "secondary photons" able to dissociate other molecular species.  We adopt the dissociation factors $p_M$ calculated by \cite{Gredel89} for various species, leading to a rate per unit volume of $\zeta(\Hr_2^+)[n(\Hr_2)/n_\Hr]n(M)p_M/(1-\omega)$ \citep{FlPi07}, where we adopt a grain albedo $\omega=0.5$. For CO, we adopt the high-abundance, high-temperature limit of $p_{\rm CO}=28$ for $\omega=0.5$ \citep{Gredel87}. Photodetachment of electrons from PAHs by secondary photons is also included \citep{FlPi03b}.

\subsubsection{X-ray heating}
The heating efficiency of the X-rays and cosmic rays is treated collectively. We use the results of \cite{DaYL99}, who studied the heating produced by high energy electrons interacting with a partly ionized gas mixture of H, H$_2$ and 10\% of He. The heating efficiency $\eta_h$ is calculated for ionization fractions $<0.1$ and includes the fraction of initial primary electron energy lost by elastic scattering with neutrals and electrons and by rotational excitation of H$_2$ (rapidly thermalized at our densities).  The total heating rate per unit volume by X-rays and cosmic rays is given by:
  \eqb
  \Gamma_\mathrm{X+CR}(z) = \eta_{h} H_X n_\Hr =
  \frac{10r\eta_{\Hr_2}+\eta_\Hr}{10r+1} \zeta(\Hr_2^+)
  \epsilon_{\mathrm{H}_2} n_\Hr \eqe
where $\epsilon_{\mathrm{H}_2}=20$~eV follows from the definition of $\zeta(\Hr_2^+)$ in eq.~\eqref{e:zetah2}, $r=n(\Hr_2)/n(\Hr)$, and $\eta_{\Hr_2},\eta_\Hr$ are respectively the heating efficiencies for the H$_2$-He and the H-He ionized mixtures, fitted as a function of the electron fraction $x_{\rm e}=n_{\rm e}/n_\Hr$ in the form
  \eqb \eta = 1+\frac{\eta_0-1}{1+ cx_{\rm e}^a } \eqe
with $\eta_0=0.055$, $c=2.170$, $a=0.366$ for H$_2$-He, and \mbox{$\eta_0=0.117$}, $c=7.950$, $a=0.678$ for H-He \citep[][Table 7]{DaYL99}.

\section{Results}\label{s:results}

  \begin{figure*}\centering
  \includegraphics[angle=270.,width=\textwidth]{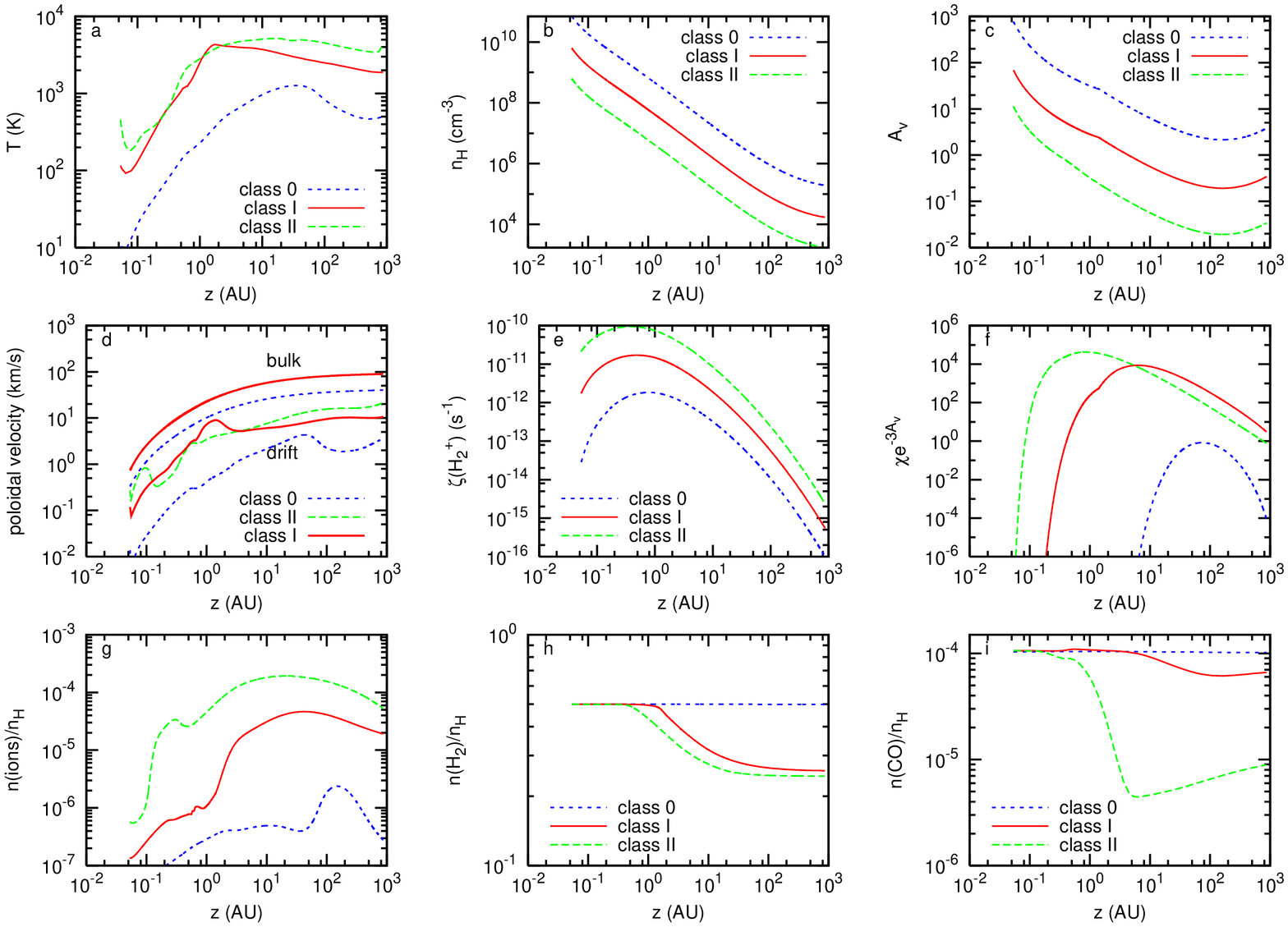}
  \caption{Various variables along a flow line anchored at 1 AU, for the dense MHD disk wind solution and three typical classes of stars: class~0 ($\Mstar=0.1\Msun$, $\Macc=5\times10^{-6}\Msun$ yr$^{-1}$; dotted blue curves), class~I ($\Mstar=0.5\Msun$, $\Macc=10^{-6}\Msun$ yr$^{-1}$; solid red curves) and active class~II ($\Mstar=0.5\Msun$, $\Macc=10^{-7}\Msun$ yr$^{-1}$; dashed green curves). (a) The gas temperature $T$; (b) the number density of hydrogen nuclei $n_\Hr$; (c) the visual extinction $A_V$ to the star, in magnitudes; (d) the poloidal components of the bulk flow speed $v$ (smooth top curves) and calculated drift speed $v_{in}$ (lower 3 curves); Note that the Class~II and Class~I have the same $\Mstar$, hence the same bulk flow speed for $r_0=1$~AU; (e) the X-ray H$_2$ ionization rate per molecule $\zeta(\Hr_2^+)$; (f) the effective radiation field in Draine units $\chi\exp^{-3A_V}$ (attenuation factor appropriate for carbon photoionisation); (g) the fractional abundance of ions; (h) the fractional abundance of $\Hr_2$ (upper limit in the Class~II model, see Section~\ref{s:H2}); (i) the fractional abundance of CO (upper limits in the Class~I/II models, see Section~\ref{s:CO}). A color version of this figure is available in the on-line edition of this journal.}
  \label{f:classes} \end{figure*}

  \begin{figure*}\centering
  \includegraphics[angle=270.,width=\textwidth]{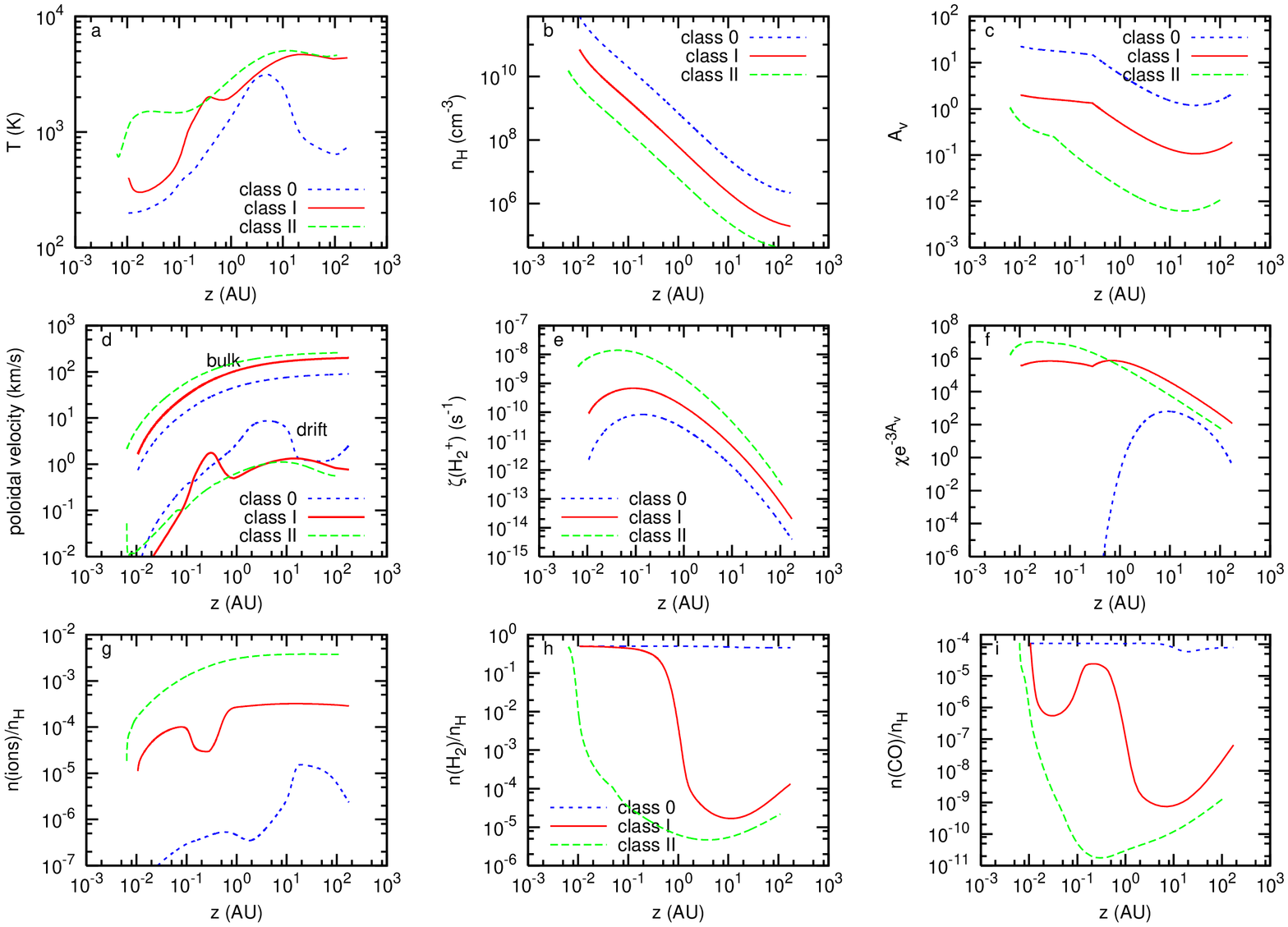}
  \caption{Same as Figure~\ref{f:classes} for streamlines anchored just beyond the sublimation radius: $r_0=\Rsub=0.2$~AU in the Class~0 and Class~I, and 0.11~AU in the Class~II.  These calculations assume no self-shielding of H$_2$ or CO (i.e.~no molecules surviving on streamlines launched inside $\Rsub$). The resulting abundances set a minimum photodissociation timescale on outer streamlines, to be compared with the flow timescale (see Figure~\ref{f:rat-c1}). A color version of this figure is available in the on-line edition of this journal.}
  \label{f:rsub-classes} \end{figure*}

To investigate how the chemical content of protostellar disk winds evolves in time with the decline of accretion rate, we considered three sets of parameters representative of the main evolutionary stages of young low-mass stars with bright jets:
\begin{itemize}
\item a very young Class~0 protostar (\mbox{$\Macc=5\times10^{-6}\Msun$ yr$^{-1}$}, \mbox{$\Mstar=0.1\Msun$}), where accretion is high and the star has not yet reached its final mass; the chosen parameter values describe e.g.~the heavily embedded sub-millimeter exciting source of the HH~212 jet \citep{lee06}.
\item a Class~I source ($\Macc=10^{-6}\Msun$ yr$^{-1}$, $\Mstar=0.5\Msun$), where the star has accumulated most of its final mass, but residual infall and accretion still proceed at a fast pace; the chosen parameters describe e.g.~the infrared exciting source of HH26 \citep{antoniucci}.
\item an active Class~II star ($\Macc=10^{-7}\Msun$ yr$^{-1}$,
$\Mstar=0.5 \Msun$), representative of optically visible T Tauri stars with bright optical microjets, e.g.~DG Tau or RW Aur \citep{Gullbring00}.
\end{itemize}
With our choice of parameters, and the self-similar scalings of the MHD solution, the wind density $n_H\propto\Macc\Mstar^{-1/2}r_0^{-1.5}$ drops by a factor 10 from one Class to the next. Our adopted stellar radius of $3\Rsun$ gives the same accretion luminosity of $5L_\odot$ in the Class~0 and Class~I models, and 10 times smaller in the Class~II, leading to a hot spot coverage fraction $\delta_\hs$ of 3.2\% and 0.32\% respectively (cf.~Section~\ref{s:UVfield}).

We adopt a thermal X-ray spectrum of characteristic energy $kT_X=4$ keV and luminosity $L_X=10^{30}$ erg/s, typical of the "hard" time-variable component in solar-mass young stars and jet-driving protostars \citep{Imanishi03, Guedel07}. $L_X$ is a time-averaged value of 30\% of the median flare luminosity in solar-mass YSOs, based on a typical flare interval of 4-6 days \citep{Wolk05}. The exact value of $kT_X$ has no major effect on our results (see Section~\ref{sec:xray-tau}).

Note that the flow crossing timescale out to the recollimation point is very short, only 50~yrs for a streamline launched at 1~AU in the Class~I model. The chemical composition of the disk wind will thus deviate substantially from published "static" disk atmosphere models such as, e.g.~\cite{Glass04,nomura}. The presence of adiabatic cooling and drag heating also introduce major differences in the thermal structure. These effects are discussed in Section~\ref{s:discussion}.

We first analyse in detail the results obtained for the 3 evolutionary classes, for a representative anchor%
  \footnote{The {\it anchor} radius $r_0$ is the radius of the magnetic
  field surface in the disk midplane. The {\it launch} radius at the slow
  magnetosonic point is 6.5\% larger --- allowing centrifugal acceleration
  of the matter loaded onto the field line.}
radius $r_0=1$~AU, and for a streamline launched just outside sublimation, $r_0\simeq\Rsub$. The effect of increasing $r_0$ is discussed later on, in Section~\ref{s:anchor}.

\subsection{Results for $r_0=1$~AU and $r_0=\Rsub$ for the 3 classes}
Figure~\ref{f:classes} shows the main physical parameters integrated as a
function of $z$ along the 1~AU streamline for the Class~0/I/II models:
temperature, ion-neutral drift speed, ionisation fraction, H$_2$ and CO
abundance. Also plotted are the (prescribed) flow density and bulk speed,
and the calculated $A_V$ to the central star, attenuated X-ray ionisation
rate of H$_2$, and "effective" FUV field at $1\,000$~{\AA} in Draine units
(e$^{-3A_V}$ being the attenuation factor for Carbon photoionisation in our
chemical network). In Figure~\ref{f:rsub-classes}, we show for comparison
the results obtained for streamlines with an anchor radius just beyond the
sublimation radius, computed assuming no H$_2$ or CO self-shielding
(i.e.~assuming no molecules on streamlines launched inside $\Rsub$).

\subsubsection{Radiation field and ionization structure}
The ionization fraction is a key parameter governing the temperature profile, through the drag heating term. It is controlled by attenuation of the radiation field through inner wind streamlines. The behavior is qualitatively similar along the streamlines launched from 1~AU and from $\Rsub$.

The A$_V$ towards the star is highest at the wind base, where the line of sight crosses the densest wind layers (see Fig.~\ref{f:model}) and drops steadily as material climbs along the streamline (Figure~\mbox{\ref{f:classes}-\ref{f:rsub-classes}c}). This causes a steep rise in the effective attenuated radiation field, until the $1/R^2$ dilution factor takes over (Fig.~\mbox{\ref{f:classes}-\ref{f:rsub-classes}e,f}). Combined with the wind density fall-off (Fig.~\mbox{\ref{f:classes}-\ref{f:rsub-classes}b}), this generates a global increase in ionisation fraction along the streamlines out to $30-100$ AU, until recombination sets in (Fig.~\ref{f:classes}-\ref{f:rsub-classes}g). In our models, ionization is dominated by stellar \mbox{X-rays} for $A_V>3$ mag, and by FUV photons further out.

As the wind density drops by a factor 10 from Class~0 to Class~I to Class~II (Fig.~\ref{f:classes}-\ref{f:rsub-classes}b), the ionisation fraction, \mbox{$X({\rm i}^+)\equiv n({\rm ions})/n_\Hr$}, increases by a factor 3 to 10. Indeed, the smaller attenuating column through the inner wind (Fig.~\ref{f:classes}-\ref{f:rsub-classes}c) leads to higher X-ray and FUV ionization rates. At the same time, the lower wind density reduces recombination. Both effects work in the same direction. 

  \begin{figure*}\centering
  \includegraphics[angle=270.,width=0.75\textwidth]{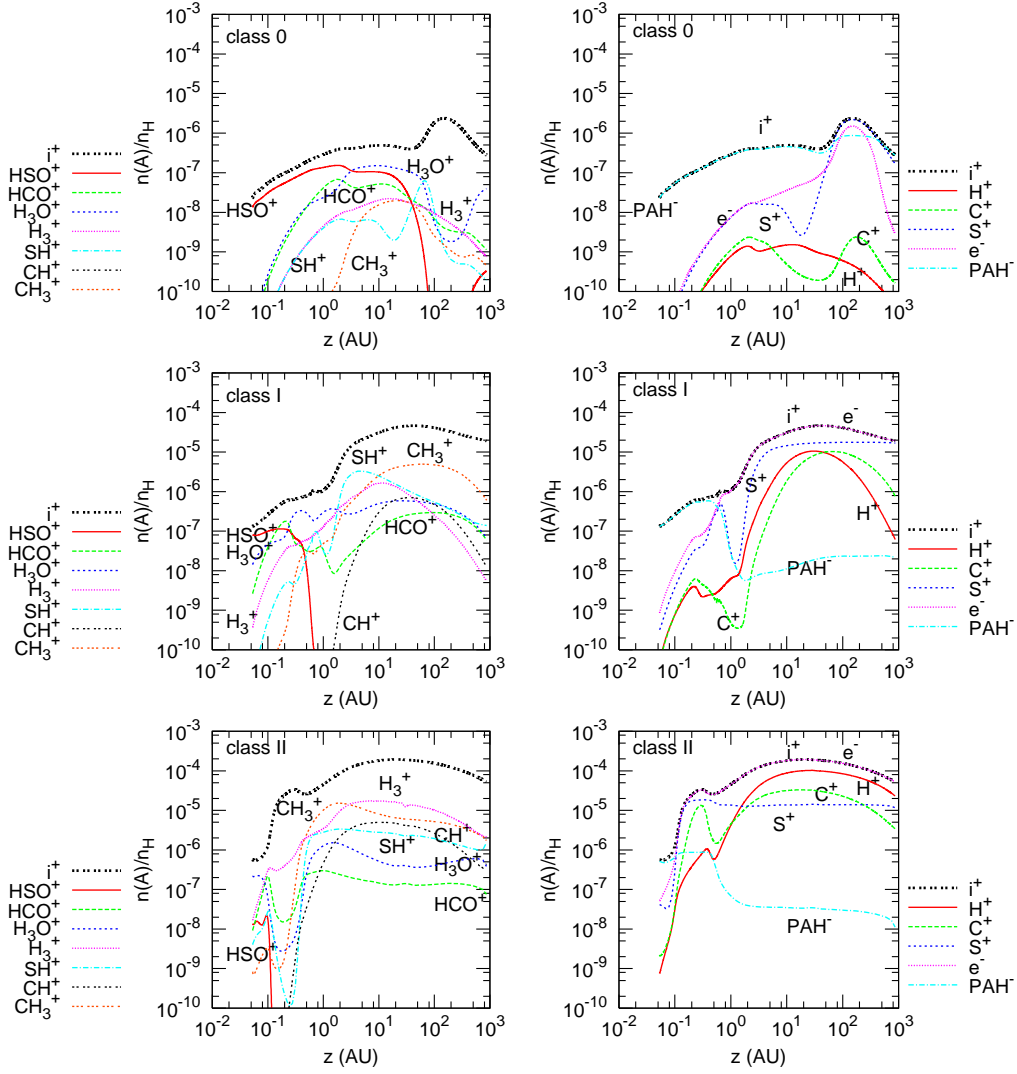} 
  \caption{The main ionization contributors along a disk wind streamline launched from 1~AU, for our Class~0, Class~I, and Class II models (top to bottom). The left panels plot the fractional abundances relative to H nuclei of molecular ions; the right-hand panels plot abundances of atomic ions and negative species (electrons and PAH$^-$). The total ionisation fraction, $X({\rm i}^+)\equiv n({\rm ions})/n_\Hr$, is also plotted as a thick black curve. A color version of this figure is available in the on-line edition of this journal.}
  \label{f:ions}\end{figure*}

  \begin{figure*}\centering
  \includegraphics[width=0.7\textwidth]{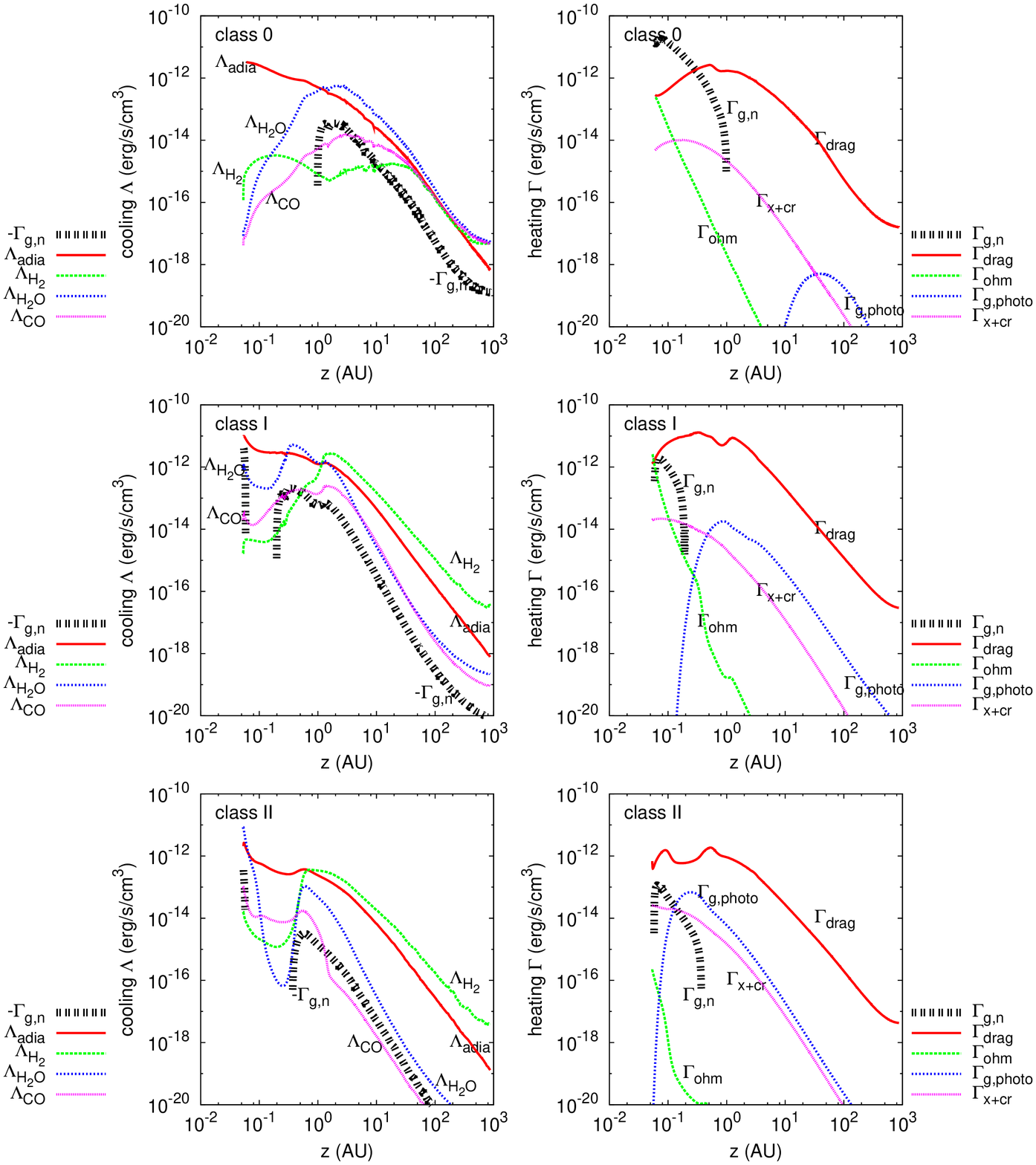}
  \caption{The main cooling (left panels) and heating (right panels) processes along a disk wind streamline launched from 1 AU, for our Class~0, Class~I, and Class~II models (top to bottom). The heat transfer from grain to gas, $\Gamma_{g,n}$, appears in the cooling curves when negative, and in the heating curves when positive. Other plotted cooling/heating terms are: $\Lambda_{\rm adia}$: adiabatic cooling; $\Lambda_{\rm H_2, H_2O, CO}$: line cooling from H$_2$, H$_2$O, CO; $\Gamma_{\rm drag}$: ambipolar diffusion heating; $\Gamma_{\rm ohm}$: ohmic heating; $\Gamma_{\rm g,photo}$: photoelectric effect on grains; $\Gamma_{\rm X,cr}$: X-ray and cosmic ray heating. OH line cooling never dominates and is not plotted for clarity. A color version of this figure is available in the on-line edition of this journal.}
  \label{f:heatcool}\end{figure*}

Fig.~\ref{f:ions} shows the main ionization contributors along the 1~AU streamline for the Class~0, Class~I, and Class~II models. From these results, it can be seen that the main contributors depend on the overall ionization degree, $X({\rm i}^+)$:
  \begin{description}
  \item[when $X({\rm i}^+)\le10^{-6}$:]The main contributors to the negative charge are the PAHs. The main positive carriers are molecular ions, with the most abundant being HSO$^+$ and H$_3$O$^+$; and H$_3^+$ (a direct product of X-ray ionization) also contributing in the Class~II.
  \item[when $X({\rm i}^+)>10^{-6}$:]The main contributors to the negative charge are the electrons. The main positive carriers are the atomic ions S$^+$, C$^+$, and H$^+$, which recombine more slowly with electrons than molecular ions do. 
  \end{description}

An interesting result of our calculations is the relatively high abundances of CH$^+$, SH$^+$, and H$^+$ of $10^{-6}- 10^{-4}$ reached along the Class~I and Class~II streamlines (Fig.~\ref{f:ions}). CH$^+$ and SH$^+$ are formed by endothermic reaction of C$^+$ and S$^+$ with H$_2$, when the wind temperature exceeds about $3\,000$~K. H$^+$ forms by X-ray ionisation and charge exchange, but also by photodissociation of CH$^+$, with carbon playing the role of a catalyst through the reaction chain:
  \eqab \Cr + {h\nu} &\rightarrow& \Cr^+ , \label{e:re1a}\\ \Cr^+ +
  \Hr_2 &\rightarrow& \Cr\Hr^+ + \Hr -4640\ \K , \label{e:re1b}\\
  \Cr\Hr^+ + {h\nu} &\rightarrow& \Cr + \Hr^+ . \label{e:re1c} \eqae
Through this repeated formation cycle, H$_2$ is partly converted into H$^+$, helping to offset H$^+$ destruction by charge exchange. The flow crossing timescale is also too short for significant H$^+$ radiative recombination. This example clearly illustrates the out-of-equilibrium and hybrid nature of the chemistry in MHD disk winds, combining elements from both C-shocks (strong heating by ion-neutral drift) and photo-dissociation regions (abundant C$^+$ and S$^+$).

\subsubsection{Drift speed, heating/cooling processes, and temperature behavior}\label{s:vdrift}
Thanks to X-ray and UV ionization, we find that the drift speed remains relatively low along the streamlines, at 10\% of the bulk poloidal speed or less, as seen in Fig.~\ref{f:classes}-\ref{f:rsub-classes}d. This validates a posteriori the single fluid approximation made in computing the MHD dynamical solution. It also keeps the drag heating and gas temperature to a moderate value. Note that the ratio of drift speed to poloidal speed scales as $v_{in}/\vp\propto\Mstar\Macc^{-1}X({\rm i}^+)^{-1}<\sigma v>_{in}^{-1}$, so that the single fluid approximation remains fulfilled in the Class~0 jet (small $\Mstar$, large $\Macc$) despite its low ionisation.

The main heating and cooling terms for $r_0=1$~AU are plotted in Fig.~\ref{f:heatcool}. Collisions with warm dust can be important near the wind base when densities exceed a few $10^{9}$~cm$^{-3}$, but ambipolar "drag" heating by ion-neutral collisions, $\Gamma_{\rm drag}$, quickly prevails as density drops. The predominance of ambipolar diffusion over other sources of heating (shocks excepted) is a widespread characteristic of steady MHD-driven  protostellar winds from low-mass sources, and results from the strong accelerating $\Jb\times\Bb$ force combined with a low wind ionisation \citep[see][]{Ruden90,Safi93a,Garcia01a}. Cooling is dominated by adiabatic expansion out to $z\simeq1$~AU, and by radiative cooling further out, due to the high molecular abundances. The main coolant is H$_2$ in the Class~I and Class~II models, and H$_2$O in the Class~0 model.

The resulting behavior of temperature along the 1~AU streamlines is qualitatively similar regardless of the accretion rate (Fig.~\ref{f:classes}a): After initial cooling due to adiabatic expansion, drag heating takes over and heats up the gas to $1\,000-5\,000$~K. This leads to a sharp increase in molecular cooling (denoted $\Lambda_{\rm rad}$ thereafter) which (assisted by collisional dissociation cooling in the Class~I/II models) eventually overcomes $\Gamma_{\rm drag}$, leading to a temperature turnover followed by a slow decline where both terms remain in close balance.
 
The balance between $\Gamma_{\rm drag}$ and $\Lambda_{\rm rad}$ on the 1~AU wind streamline is maintained thanks to the powerful thermostatic behavior of molecular cooling, which is highly sensitive to temperature. Any slight excess/deficit of drag heating leads to an increase/decrease in radiative cooling until the two terms match again. In the outer wind, radiative cooling approaches the low-density limit (i.e.~$\Lambda_{\rm rad}\simeq n_\Hr^2f_{\rm rad}(T)$), therefore the asymptotic jet temperature profile is determined by the condition $f_{\rm rad}(T)\simeq\Gamma_{\rm drag}/n_\Hr^2$. The latter term slowly declines with distance in our models, leading to the slowly declining temperature. Furthermore, the self-similar scalings for $\vec{J}, \vec{B}, n_\Hr$ in eq.~(9) of \cite{Garcia01a} yield a scaling for \mbox{$\Gamma_{\rm drag}/n_\Hr^2\propto\Mstar^3\Macc^{-2}r_0^{-1}X({\rm i}^+)^{-1}$}, explaining why we find much cooler jets in the Class~0 case. Final temperatures are only $\sim500-700$~K, compared to $2\,000$~K in the Class~I, and $3\,000$~K in the Class~II models.

The streamlines from $r_0=\Rsub$ in the Class~I and II cases show a slightly different asymptotic temperature behavior, with a flatter isothermal "plateau" at large distances (Fig.~\ref{f:rsub-classes}a). Molecules are heavily photodissociated on these two streamlines, so that ambipolar diffusion heating is balanced by adiabatic cooling alone. The flat temperature "plateau" resulting from this balance between $\Gamma_{\rm drag}$ and $\Lambda_{\rm adia}$ is a well known typical property of {\it atomic} self-similar MHD disk winds \citep{Safi93a,Garcia01a}. The lower atomic plateau temperature $\simeq4\,000$~K found here, compared to $\simeq10^4$~K in \citet{Garcia01a}, results from our different adopted MHD wind solution, which is typically 2 times slower and 5 times denser (see Section~\ref{s:dynamics}).

It is also noteworthy that, for both values of $r_0$, the initial temperature {\it rise} is much slower in the Class~0 jet; as shown in \cite{Garcia01a} (equations 24 and 26), the slope of the initial temperature rise, where $\Lambda_{\rm adia}$ dominates the cooling, is determined by the {\it ratio} of $\Gamma_{\rm drag}$ to $\Lambda_{\rm adia}$, which scales as $\Mstar^2/(\Macc r_0)$ in self-similar disk winds for a fixed ionisation fraction. This scaling is smaller in our Class~0 than in our Class~I case by a factor $5^3=125$, explaining the delay to heat the gas in the Class~0 jet despite a 10 times lower ionisation fraction than in the Class~I jet.

\subsubsection{H$_2$ survival}\label{s:H2}
The calculated H$_2$ abundances along streamlines launched from 1~AU and from $\Rsub$ are plotted in Fig.~\ref{f:classes}-\ref{f:rsub-classes}h, for the 3 classes of sources. Below, we first analyse the results in detail for the Class~I case, and then discuss the Class~II and Class~0 cases.

  \begin{figure}\centering
  \includegraphics[width=0.5\textwidth]{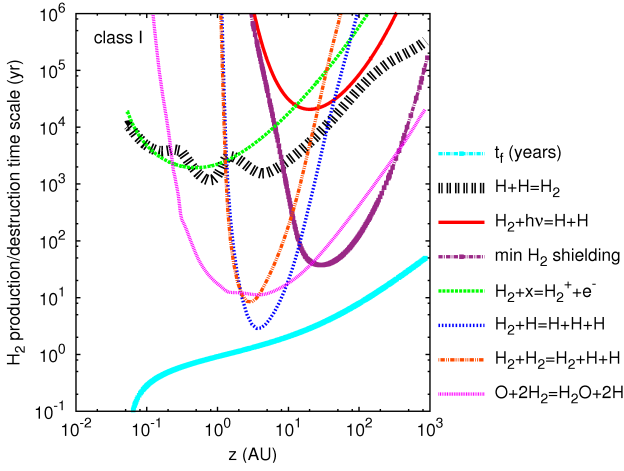}
  \caption{The net timescales (after subtraction of any reverse reaction rate) of the principle reactions that form or destruct H$_2$ are plotted versus altitude along a streamline launched at 1 AU for our Class~I model. The flow crossing timescale $t_{\rm f}$ is also plotted at the bottom for comparison. Thin curves denote reactions whose net effect leads to destruction, and thick curves those that lead to net creation. The reactions are $\mathrm{H+H=H_2}$: catalytic formation of H$_2$ on grains; $\mathrm{H_2+h\nu=H+H}$: photodissociation (with the minimum self-shielded rate shown on a separate curve); $\mathrm{H_2+x=H_2^++e^-}$: \mbox{X-ray} ionisation; $\mathrm{O+2H_2=H_2O+2H}$: combined oxydation reactions with O and OH; $\mathrm{H_2+H}$ and $\mathrm{H_2+H_2}$: collisional dissociation. The latter becomes very efficient near $z\simeq2$~AU, where the wind temperature reaches $4\,000$~K. A color version of this figure is available in the on-line edition of this journal.}
  \label{f:rat-c1} \end{figure}

\paragraph{Class~I case:}
The streamline launched just outside \mbox{$\Rsub=0.2$}~AU suffers heavy H$_2$ destruction (see Fig.~\ref{f:rsub-classes}h). Self-shielding is not operative (we assumed no molecules on inner, dust-free streamlines) and the FUV field is very strong. Hence most H$_2$ molecules are quickly photodissociated at $z\simeq1$~AU. Balance with reformation on dust yields a final H$_2$ abundance $\simeq10^{-4}$.

In contrast, half of the H$_2$ molecules are found to survive along the 1~AU streamline (see Fig.~\ref{f:classes}h). This may be understood by examining the timescales of the main formation/destruction processes of H$_2$ along this streamline, plotted in Fig.~\ref{f:rat-c1}. The flow timescale up to the current point, $t_{\rm f}$, is also plotted for comparison and is seen to be very short (50 yrs out to $1\,000$~AU). With our assumption of a mean H$_2$ abundance of half the local value, the self-shielded H$_2$ photodissociation timescale (red curve in Fig.~\ref{f:rat-c1}) is now much longer than the flow timescale. As a conservative check of this conclusion, we also plot in Fig.~\ref{f:rat-c1} (thick pink curve with symbols) the {\it minimum} H$_2$ photodissociation timescale, where $N(\Hr_2)$ and the corresponding self-shielding factor are calculated assuming an H$_2$ abundance on inner streamlines equal to its minimum value, as obtained on the streamline $r_0=\Rsub$. Even in this overly pessimistic case, the minimum H$_2$ photodissociation timescale remains an order of magnitude longer than the flow time at all $z$ (Fig.~\ref{f:rat-c1}), confirming that H$_2$ will escape photodissociation on the 1~AU streamline in the Class~I model.

Only two processes are found to be fast enough to affect the H$_2$ abundance on the 1 AU streamline: The first to occur is endothermic neutral-neutral reactions with O and OH to form H$_2$O and H atoms (denoted collectively as $\mathrm{O+2H_2=H_2O+2H}$ in Fig.~\ref{f:rat-c1}):
  \eqab \Or + \Hr_2 &\rightarrow& \Or\Hr + \Hr-2980\ \K\label{e:re2a}\\
  \Or\Hr + \Hr_2 &\rightarrow& \Hr_2\Or + \Hr-1490\ \K\label{e:re2b}. \eqae
The second process is collisional dissociation by H and H$_2$. The timescale shortens by 6 orders of magnitude as temperature increases from $2\,000$~K to $4\,000$~K, and approaches the flow time around $z\simeq2$~AU, so that half of H$_2$ is destroyed by 10~AU. Further out, collisional dissociation shuts off due to both the temperature decline and the density fall-off; the H$_2$ abundance thus remains "frozen-in" at half its initial value (Fig.~\ref{f:classes}h). We conclude that streamlines will turn from mostly atomic to mostly molecular around $r_0 \simeq 1$~AU for our Class~I model.

\paragraph{Active Class~II case:}
The H$_2$ abundance shows a similar behavior to the Class~I case. The streamline from $\Rsub$ suffers heavy photodissociation, with a final abundance of $\simeq2\times10^{-5}$ (Fig.~\ref{f:rsub-classes}h), while half of the molecules survive on the 1~AU streamline (Fig.~\ref{f:classes}h). However, the {\it minimum} photodissociation timescale on the 1~AU streamline, computed with the same method as above, is now shorter than the flow time around $z\simeq30$~AU; therefore, our local approximation to self-shielding might significantly underestimate H$_2$ photodestruction on the Class~II streamline launched at 1~AU. Our H$_2$ abundance in this particular case should thus be viewed conservatively as an upper limit, pending more detailed non-local calculations.

\paragraph{Class~0 case:}
Here, wind densities are so high that dust screening against photodissociation is very effective, even on the streamline just outside $\Rsub$. In addition, temperature is so low that oxydation reactions and collisional dissociation are slower than the flow crossing time, except briefly on the $\Rsub$ streamline. Therefore, the Class~0 disk wind retains $90\%-100\%$ of its full initial H$_2$ content on dusty streamlines launched from $r_0\ge\Rsub=0.2$~AU.

\subsubsection{Main O, C, and S-bearing species along the 1 AU streamline}\label{s:CO}

  \begin{figure*}\centering
  \includegraphics[width=\textwidth]{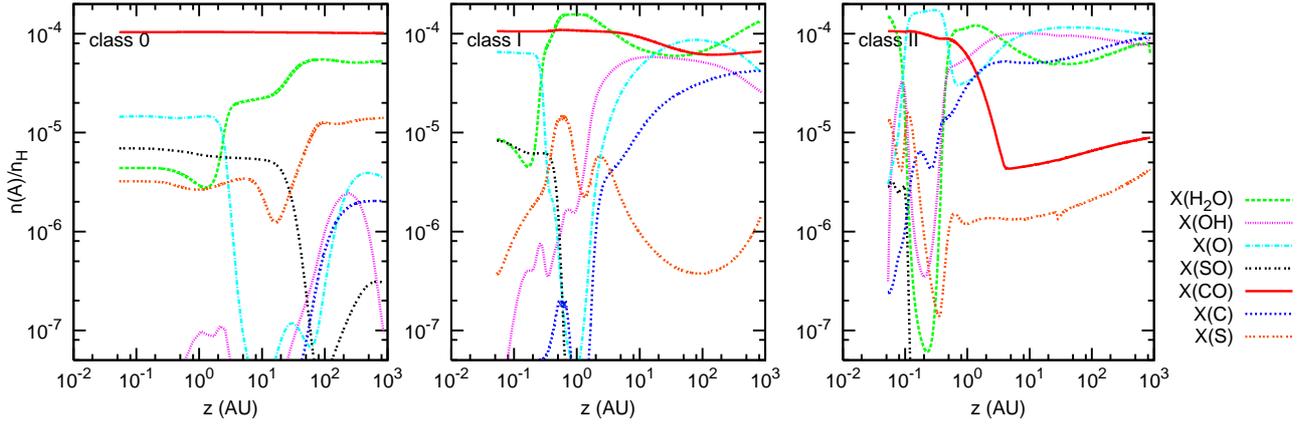}
  \caption{Abundances relative to H nuclei of important O, C, and S-bearing species along a flow line launched at 1~AU, for our Class~0 (left), Class~I (middle) and Class~II models (right). A color version of this figure is available in the on-line edition of this journal.}
  \label{f:ab-cl}
  \end{figure*}
The fractional abundances of the main O- and C-bearing neutral species with respect to H nuclei are plotted in Fig.~\ref{f:ab-cl} along the 1~AU streamline for the 3 classes of sources. S and SO are also plotted, as they were identified recently as tracers of molecular jets in Class~0 sources \citep{Dio09, Dio10,tafalla10}.
  
\paragraph{CO:}
We find that no CO destruction occurs on the Class~0 streamline from 1~AU,
thanks to the very efficient dust shielding provided by dense inner wind
layers. In contrast, CO is partly destroyed by FUV photodissociation along
the Class~I/II streamlines. Once the FUV field drops down, CO reforms mainly
through the reaction $\mathrm{HCO^++e^-}\rightarrow\mathrm{CO+H}$ but the
timescale is longer than the flow time and the final abundance per H nucleus
is only $5\times10^{-5}$ in the Class~I jet, and $10^{-5}$ in the Class~II
jet. These latter values should be considered conservatively as upper
limits, as CO self-shielding could be significantly overestimated by our
approximate "local" treatment (see Section~\ref{s:anchor} below).
We thus predict an abundance ratio of H$_2$/CO $\ge 0.5 \times 10^4$ in all cases.

\paragraph{H$_2$O:}
When the wind temperature exceeds $\simeq400$~K, oxygen not locked in CO is converted to H$_2$O through the endothermic reactions of O and OH with H$_2$ in eq.~\eqref{e:re2a}-\eqref{e:re2b}. Together with H$_3$O$^+$ recombination, this reaction is efficient enough to balance the water destruction processes active in Class~I/II jets (FUV photodissociation, photoionisation, reaction with H$_3^+$). The asymptotic abundance of water is thus similar and quite high on the 1~AU streamline for all 3 classes%
  \footnote{In the Class~0 streamline, cold dust grains hold an additional
  H$_2$O reservoir in the form of ice, of assumed abundance $10^{-4}$
  \citep{FlPi03b}, that could be later released in the gas phase in shock
  waves.},
at $\simeq5-10\times10^{-5}$. 

\paragraph{C, O, S atoms, OH and SO:}
OH and atomic C and O have low abundances along the Class~0 streamline, where CO and H$_2$O are well shielded. In contrast, high abundances of these species $\simeq0.5-1\times10^{-4}$ are predicted in the outer regions of the Class~I and Class~II streamlines, following CO and H$_2$O partial photodissociation (see Fig.~\ref{f:ab-cl}). Concerning sulfur, it is mainly in the form of atomic S beyond 100~AU along the Class~0 streamline, and is mainly photoionized into S$^+$ in the Class~I/II cases, where the FUV field is more intense (see Fig.~\ref{f:ions}). The asymptotic abundance of SO is substantial in the Class~0 only, at a level of $3\times10^{-7}$. These predicted characteristics for Class~0 jets are in line with the relative abundance of SO to CO $\simeq2\times10^{-3}$ reported by \citet{tafalla10}, and with the mass-flux traced by atomic sulfur lines being possibly as large as that inferred from CO \citep{Dio10}.

\subsection{Effect of increasing launch radius in the Class~I model}
\label{s:anchor}
\begin{figure*}\centering
  \includegraphics[width=\textwidth]{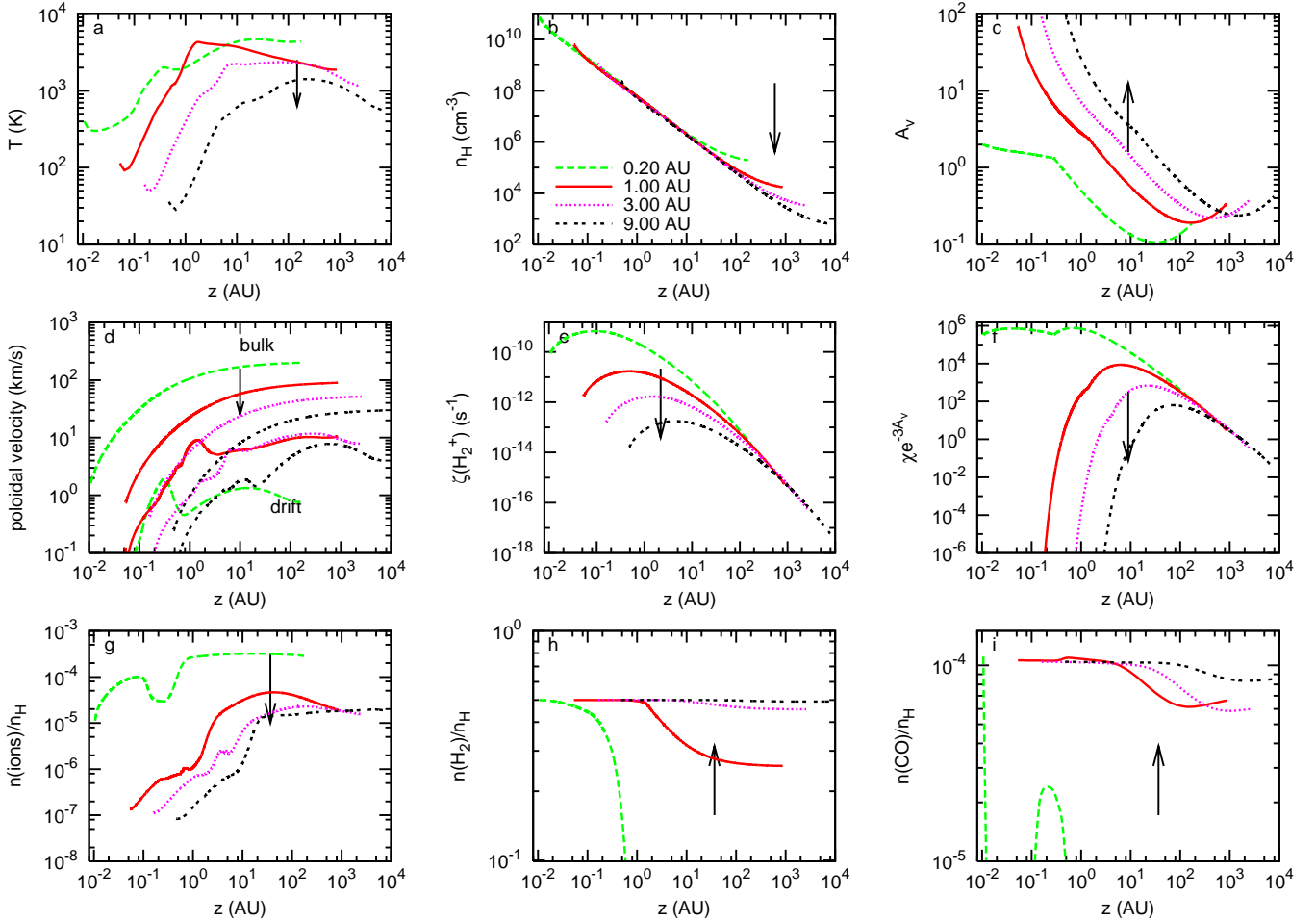}
  \caption{Thermo-chemical structure for a range of streamlines in our Class~I jet: $\Rsub\simeq0.2$~AU(green dashed, taken from Fig.~\ref{f:rsub-classes}), Class I jet: 0.34~AU (blue dotted), 0.58~AU (green solid), 1~AU (red solid), 3~AU (pink dotted), and 9~AU (black dashed). The arrow shows the effect of increasing launch radius on the plotted quantities, which are the same as in Fig.~\ref{f:classes}. CO abundances are upper limits, except on the $r_0=\Rsub$ streamline (calculated with no self-shielding). A color version of this figure is available in the on-line edition of this journal.}
  \label{f:anchor}\end{figure*}

In Fig.~\ref{f:anchor} we illustrate in the Class~I case the effect of the launch radius on the wind chemistry and temperature. We plot a range of $r_0$ of $1-9$~AU corresponding to a range in final flow speed of $90~\kms$ to $30~\kms$. The streamline from $\Rsub=0.2$~AU is also shown for comparison. The main effect is that the flow has an "onion-like" thermal-chemical structure, with streamlines launched from larger radii having higher H$_2$ abundance and lower temperature and ionization. The behavior with $r_0$ is discussed in more detail below.

We do not present results for the interval $0.2<r_0<1$~AU in the Class~I jet, as the minimum H$_2$ photodissociation timescale there becomes similar to or less than the flow time (see discussion in Section~\ref{s:H2}). Accurate H$_2$ abundances on these inner streamlines await a detailed non-local treatment of H$_2$ shielding, not taken into account in the present preliminary approach. For the same reason, we do not present results at $r_0>1$~AU for the Class~II jet, but we note that the effect of increasing $r_0$ will be qualitatively the same as in the Class~I case (i.e.~higher H$_2$ abundance and lower temperature). We also do not present results at $r_0>1$~AU in the Class~0 case, since most H$_2$ and CO molecules survive already at a launch radius just beyond the sublimation radius $\Rsub=0.2$~AU (see Section~\ref{s:H2}).

\subsubsection{Ionisation and drift speed at various $r_0$}\label{s:tempr0}
Wind streamlines are initially less ionised at larger $r_0$ (Fig.~\ref{f:anchor}f,g), due to the larger attenuation of X-rays and FUV photons by intervening streamlines (Fig.~\ref{f:anchor}c). At large distances, they recombine to an asymptotic fractional ionisation of a few $10^{-5}$, dominated by S$^+$.

The drift speed does not exceed about $10~\kms$ and remains less than 30\% of the bulk flow speed out to $r_0=9$~AU (Fig.~\ref{f:anchor}d). Therefore, magnetocentrifugal forces appear able to lift molecules out to disk radii of 9~AU for the typical parameters of Class~I sources, in the MHD solution investigated here.

\subsubsection{Temperature and molecule survival at various $r_0$}
\label{s:chemr0}
The gas temperature at larger launch radii follows a similar behavior to that found at $r_0=1$~AU (i.e.~a gradual rise on a scale $z\simeq r_0$, followed by a shallow decrease), but the temperature peak is shifted to larger $z$ (i.e.~lower $n_\Hr$) and reaches a smaller peak value (see Fig.~\ref{f:anchor}a). Therefore, H$_2$ chemical oxydation and collisional dissociation are less efficient than for $r_0=1$~AU. As a result, the H$_2$ abundance increases with $r_0$ and more than 90\% of the molecules survive for launch radii $\ge3$~AU (Fig.~\ref{f:anchor}h). Photodissociation of H$_2$, which was already too slow for $r_0=1$~AU, is further reduced by the additional screening from intervening wind streamlines and plays no role.

In contrast, we find that photodissociation remains the major destruction process for CO out to $r_0=9$~AU. The drop in CO abundance towards the inner $r_0=\Rsub$ streamline largely exceeds the factor 2 assumed in our "local" computation of the CO self-screening column (see Fig.~\ref{f:anchor}i). The plotted CO abundances for $r_0>\Rsub$ should thus be considered as upper limits only, and a detailed non-local treatment of self-shielding will be needed to obtain more accurate values. 

\section{Discussion}\label{s:discussion}

\subsection{Comparison with static disk atmosphere models}
It is instructive to compare our results with thermo-chemical calculations
of irradiated {\it static} molecular disks at a radius of 1~AU.
\citet{Glass04} focussed on X-ray disk irradiation, with a similar X-ray
flux to ours ($L_X=2\times10^{30}$~erg~s$^{-1}$, $kT_X=1$~keV).
\citet{nomura} studied FUV irradiation by the star with standard ISM dust
properties as adopted here, and a mean unshielded flux at a distance of
140~AU of $G_0\simeq1\,000$ times the mean interstellar flux, averaged over
the interval $910-2\,000$~{\AA} (see their Fig.~4), similar to our Class~I
case (see Fig.~\ref{f:classes}f).

We find that the thermo-chemical structure of MHD disk winds launched from 1~AU in Class~I sources markedly differs from irradiated static disks at the same radius: H$_2$ survives to much greater altitudes and the gas temperature rises more gradually. We briefly discuss and explain these differences below.

\subsubsection{H$_2$ survival and wind shielding}
In static disk models at 1~AU, H$_2$ is destroyed by X-rays or FUV photons above $z\simeq0.1$~AU \citep{Glass04,nomura}. In contrast, on the 1~AU streamline of the Class~I disk wind, half of the gas remains molecular beyond $z\simeq10$~AU and until the end of the streamline (Fig.~\ref{f:classes}h). The survival of H$_2$ against photodissociation arises from several effects:
  \begin{enumerate}
  \item The first key effect is the enhanced screening provided by dense inner wind streamlines compared to a hydrostatic flared disk geometry. As shown in Fig.~\ref{f:model}, the self-similar MHD disk wind solution chosen here produces density contours that are horizontal just above the disk surface. In contrast, density contours in a hydrostatic disk are strongly flared out, allowing a more direct illumination of the disk surface by the central star (see Fig.~1 of \citealt{nomura}). This difference is illustrated by the large $A_V$ values towards the central star in Fig.~\ref{f:classes}c; in the Class~I streamline at 1~AU, the "effective" dust-attenuated FUV field at $z\simeq0.3$~AU is $\chi\exp^{-3A_V}\simeq2$ compared to $G_0\simeq10^7$ in the static disk of \citet{nomura} at the same $(r,z)$. The peak X-ray ionisation rate is $\zeta(\Hr_2^+)\simeq2\times10^{-11}$~s$^{-1}$, instead of $6\times10^{-9}$~s$^{-1}$ at the top of the disk atmosphere in \citet{Glass04}. 
  \item The second key factor is the very short flow timescale; only 50 years from the slow point out to $1\,000$~AU for the 1~AU streamline. Static disk models assume chemical equilibrium, so that H$_2$ disappears wherever destruction reactions are faster than reformation. In disk winds, H$_2$ can still survive this situation as long as the destruction time is longer than the flow time (see Fig.~\ref{f:rat-c1}).
  \end{enumerate}

\subsubsection{Gas temperature profile}
In static disk models at 1~AU, FUV photoelectric heating or \mbox{X-ray}
heating induce a sharp increase in gas temperature up to $1\,500-5\,000$~K
within a few scale heights, at $z/r_0\simeq0.15-0.2$ \citep{nomura,Glass04}.
In our disk wind model, the rise in temperature is much more gradual, with
temperatures of a few thousand K reached only around $z/r_0\simeq1$. This
difference stems both from the greater attenuation of \mbox{X-rays} and FUV photons by inner wind streamlines, and from the powerful {\it adiabatic expansion cooling} experienced by the wind after launching. The latter term largely exceeds FUV and X-ray heating and is only slightly offset by drag heating (see Fig.~\ref{f:heatcool}), yielding a slow net heating rate. A related consequence is that temperatures above $2\,000$~K are reached only in wind regions of low density so that H$_2$ collisional dissociation and chemical oxydation is limited.

\subsection{Limitations of the model and planned improvements}
The results presented here are only a preliminary attempt at addressing the highly complex problem of the coupled ionisation, chemical, thermal, and dynamical structure of MHD disk winds from young stars. Several limitations have been introduced that we discuss below with our planned improvements:
  \begin{enumerate}
  \item We ignored feed-back of the thermal structure on the dynamics. This is in fact well justified because thermal pressure gradients are already negligible above the slow point where our integration starts, and are only dynamically important in the mass-loading region located deeper in the disk, around $z\simeq h$ (see \citealt{CaFe00}). Therefore, any deviations between our computed wind temperature and the prescription adopted by \cite{CaFe00} to compute the dynamical solution do not alter the self-consistency of the model. Nevertheless, we show in Appendix~\ref{app:heating} that the two temperatures fall close to each other near the wind base. The existence of such "slow" warm MHD solutions might then be a natural outcome of the heating mechanisms present near the surface of an irradiated molecular disk. Thermal calculations extending into the optically thick and resistive mass-loading region are under way to test this conjecture (Garcia et al., in preparation).
  \item We also ignored feed-back of the ion-neutral drift on the flow
  dynamics; we showed that this single fluid approximation is well fulfilled
  at launch radii $\le9$~AU but tends to degrade at larger $r_0$ or small
  $\Macc$, due to the scaling of
  \mbox{$\vin/\vp\propto X({\rm i}^+)^{-1}\Macc^{-1}$}. Therefore,
  ion-neutral drift may be one of the factors (together with the available
  magnetic flux) limiting the outer radial extension of molecular MHD disk
  winds in young stars. The drift will also decrease rotation signatures in
  neutral jet tracers at large $r_0$ compared to the single fluid
  prediction. Multi-fluid MHD calculations would be useful to further
  quantify this effect.
  \item We adopted a simplified treatment of shielding against photodissociation. An upcoming improvement of our model will be the inclusion of a non-local treatment of the self-screening, in order to obtain more accurate predictions of CO and/or H$_2$ abundances in the Class~I and Class~II jets.

  Shielding against photodissociation also depends on the assumed composition of inner wind streamlines launched between $\rin=0.07$~AU and $\Rsub$. As done in \citet{Ruden90} and \citet{Garcia01a}, we assumed that all wind material with $R(\theta)>\Rsub$ contributes to the $A_V$, i.e.~that dust reforms efficiently on these dense streamlines after they exit the sublimation surface. This may be too optimistic. On the other hand, when computing the minimum H$_2$ abundance at $r_0=\Rsub$, we assumed that no molecules are present on more inner streamlines to provide self-shielding, which is pessimistic. Indeed, CO overtone infrared line profiles in Class~II disks indicate the presence of rotating CO down to $\rin\simeq0.04$~AU \citep{najita03}. Another planned improvement in our model will thus be to perform thermal-chemical modelling of dust-free streamlines launched inside $\Rsub$, including H$_2$ gas phase formation processes through ion chemistry and three-body reactions, and advection from outer radii. 
  \item The dust size distribution and dust to gas ratio were assumed to be the same as in the ISM. This is probably adequate for the young Class 0 and Class~I, where the disk is young and still being fed by an infalling envelope, but there is evidence for both dust growth and dust settling starting to occur during the Class~II phase, where the envelope has dissipated. This will decrease the efficiency of dust screening against photodissociation. On the other hand, grain growth may lead to a smaller dust sublimation radius, so that more molecular streamlines can contribute to self-shielding. Knowledge of the dust size stratification in turbulent magnetized disks will be needed to assess the net effect on H$_2$ survival in MHD disk winds from evolved Class~II stars.
  \item The present exploratory work considered only one particular steady MHD disk wind solution, with a magnetic lever arm parameter $\lambda\simeq14$ best reproducing current tentative rotation signatures in atomic jets from Class~II (see Section~\ref{s:dynamics}). If jet rotation has been overestimated, slower models with smaller values of $\lambda$ would have to be considered. These denser winds (higher ejection efficiency $\xi$) would be better self-screened, and thus less subject to photodissociation. On the other hand, the effect on H$_2$ collisional dissociation is difficult to predict without detailed temperature calculations: the smaller accelerating Lorentz force per particle will tend to decrease ambipolar heating, while the lower ionisation (due to enhanced screening) will tend to increase it (eq.~\ref{eq:gdrag}). Such solutions are being developed and will be explored in forthcoming papers.
\end{enumerate}

\subsection{Predicted observational trends and future tests}
Our preliminary results on the thermo-chemistry of a dusty centrifugal MHD disk wind exhibit general properties that seem promising to explain several observed trends in the molecular counterparts of stellar jets. They are summarized below, and future observational tests are outlined.

Concerning low-mass Class~0 sources, the model predicts that dusty flow
streamlines will keep most of their H$_2$ and CO content at least down to
the sublimation radius of $\simeq0.2$~AU, i.e.~up to flow speeds of
$\simeq100~\kms$ for $\Mstar=0.1\Msun$. This seems promising to explain the
frequent detection of collimated fast H$_2$ and CO in Class~0 jets, as well
as the high ratio of H$_2$ to CO $\simeq10^4$ recently estimated in one of
them, HH211 \citep{Dio10}. Furthermore, tentative rotation signatures
reported so far in Class~0 jets \citep{lee-hh211rot,lee-hh212rot} seem
consistent with steady MHD centrifugal launching from the expected range of
disk radii $r_0\simeq0.15-0.6$~AU, when the most likely poloidal speeds are
adopted%
  \footnote{The launch radius of a centrifugally-driven MHD disk wind, as
  inferred from jet rotation signatures, scales with the poloidal speed
  approximately as $r_0\propto\vp^{-4/3}$ \citep{anderson03}.
  \cite{lee-hh211rot,lee-hh212rot} assumed a range $\vp\simeq100-200~\kms$
  to estimate launch radii in the HH211 and HH212 Class~0 jets. The smaller
  $\vp$ value seems more probable in HH211 given the very low-mass of the
  central source ($0.065\Msun$), and yields $r_0\simeq0.15$~AU. In HH212,
  rotation signatures were seen only at low radial velocities
  $|V_{\rm r}|\simeq1.5-4~\kms$ while SiO jet emission is detected up to
  $|V_{\rm r,max}|\simeq9-12~\kms$ \citep{lee-hh212rot}. The maximum radial
  velocities correspond very well to the proper motion $\simeq150~\kms$ of
  H$_2$ knots in HH212, for the estimated jet inclination of $\simeq4^\circ$
  to the plane of the sky \citep{Cod07,claussen}. Hence, the slow SiO
  rotating gas probably has a lower poloidal speed $\vp\simeq20-60~\kms$,
  yielding $r_0\simeq0.6$ AU in both jet lobes \citep{Cab09} instead of
  $0.3-0.05$ AU as estimated by \citet{lee-hh212rot}.}.
Finally, if the MHD disk wind operates over a radial extent from
$\rin=0.2$~AU to $\rout$ of $1-9$~AU, the ratio of molecular mass-flux to
accretion rate would be \mbox{$\xi\log(\rout/\rin)\simeq0.06-0.15$}, in the
same range as observations \citep{Lee07}.

As the accretion rate drops and the stellar mass increases, our modelling indicates that the MHD disk wind is more irradiated (less screening against FUV radiation) and hotter (self-similar scaling of $\Gamma_{\rm drag}$ with $\Mstar$ and $\Macc$). Therefore, the molecular H$_2$ zone moves to larger launch radii, around $r_0\ge1$~AU in our Class~I source, with a typical terminal speed of $\simeq100~\kms$ (cf.~eq.~\ref{eq:vp-r0}). The molecular region will move even further out in Class~II sources. This predicted trend is in line with possible rotation signatures reported so far in molecular Class~I/II jets (HH26 and CB26), which do appear to suggest larger magneto-centrifugal launch radii $\simeq3-10$~AU than in Class~0 jets, although statistics are admittedly still limited \citep{h2-jetrot,cb26,Cab09}. This prediction also seems promising to explain the trend towards slow, wide molecular counterparts in older Class~II jet sources. The striking example of the broad slow H$_2$ wind encompassing the HL Tau and DG Tau atomic jets, and the slow CO conical flow surrounding the atomic jet in HH~30, both suggest a "hollow" molecular wind structure consistent with this idea \citep{tak-hltau, beck08, hh30-flow}.

A third promising trend is that gas temperature on molecular streamlines is predicted to increase from $T\simeq700$~K in Class~0 jets to $T\simeq2\,000-3\,000$~K in Class~I/II jets. This is in good agreement with temperatures $\simeq300-1\,000$~K recently inferred in Class~0 jets from H$_2$ pure rotational lines studied with {\it Spitzer} \citep{Dio09,Dio10}, and $\simeq2\,500$~K in Class~I/II jets from H$_2$ ro-vibrational lines \citep{tak06,beck08}. As rovibrational lines might have a greater contribution from shock-heated gas, H$_2$ temperature estimates in Class~II jets from pure rotational lines would be useful to confirm this result.

While these trends are encouraging, more specific tests need to be carried out to validate the applicability of centrifugal MHD disk winds to molecular protostellar jets. Closer comparison to observations await more detailed calculations of self-shielding for the Class~I/II, and computations of synthetic maps and spectra in various tracers, taking into account beam smearing and NLTE effects (Yvart et al., in preparation). Several aspects promise to be particularly discriminant:

A first important test would be a confirmation of rotation in molecular jets, with a resolved spatial pattern consistent with centrifugal acceleration from the disk. Current measurements are affected by beam smearing, which can significantly lower the observed rotation signature compared to the actual underlying rotation curve \citep{Pes04}. This effect could be important in the HH212 and HH211 Class~0 jets in Orion-Perseus, whose widths $\leq0.2~\arcsec$ are smaller than the best resolution of current submm interferometers \citep{Cab07,lee-hh211rot}. The lack of resolution also makes it more difficult to recognize possible contamination of rotation signatures by jet precession, orbital motions, or shock asymmetries (see e.g.~\citealt{cerqueira,correia} for discussions of possible artefacts). The ALMA interferometer will be essential to progress on this issue.

A second prediction of the MHD disk wind model considered here is that, beyond the dust sublimation radius, the wind base is substantially shielded from stellar photons and the gas heats up only gradually through ambipolar diffusion (see Section~\ref{s:vdrift}). CO rovibrational lines, excited around $1\,000$~K, would thus be formed higher up in the atmosphere (at $z\simeq1$~AU) than in static disks heated mostly by irradiation. Comparison of predicted CO rovibrational line profiles with observed ones in Class~I/II sources in the near-infrared \citep{najita03,class2-coprofiles} may thus provide an interesting test. The high predicted abundance of $\mathrm{H_2O}>10^{-5}$ in warm MHD disk winds ejected from $\simeq1$~AU, at all evolutionary stages (see Fig.~\ref{f:ab-cl}) is also an important characteristic that may be testable with {\it Herschel} or infrared observations \citep[e.g.,][]{lars10, water-profile}.

Finally, we note that the wind thermo-chemical properties may be locally modified by internal shock waves forming in the jet due to time-variability or instabilities; which molecules are destroyed or reformed will depend on whether the shock is of C ("continuous") or J ("jump") type, which in turn depends crucially on the ionization fraction, magnetic field intensity, and H$_2$ fraction in the preshock gas \citep{lebourlot02,FlPi03b}. The present calculations provide for the first time the appropriate shock "initial conditions" in an MHD disk wind, setting the stage to calculate the predicted thermo-chemistry in internal shocks, and to compare it with shock observations. Although such developments are beyond the scope of the present paper, we note that SiO and CH$_3$OH, observed to be enhanced by 2 to 4 orders of magnitude in Class~0 jet knots \citep{Bach91,Cab07,tafalla10}, could in principle be released in a dusty disk wind following shock erosion or vaporization of grains \citep{Gus08b,guillet-jtype,guillet-ctype}.

\section{Conclusions}\label{s:concl}
We have investigated the non-equilibrium thermal-chemical structure of dusty streamlines in a "slow" self-similar MHD disk wind compatible with current observational constraints in atomic T Tauri microjets. We considered a range of accretion rates and stellar masses representative of the Class~0, Class~I, and early Class~II stages of low-mass star formation, and probing a decrease of 2 orders of magnitude in wind density. A detailed chemical network, a complete set of heating/cooling terms, and the effect of irradiation by stellar X-rays and FUV photons, were considered. Our main conclusions are the following:

\begin{itemize}
\item The MHD disk wind has an "onion-like" thermal-chemical structure, with temperature, ionization, and radiation field all decreasing as the launch radius of the streamline increases. The wind is sufficiently ionized to accelerate neutrals out to disk radii of $\simeq9$~AU, but is sufficiently self-screened and cool to remain molecular on streamlines launched beyond some minimum radius $r_0$. For the MHD solution explored here, this radius is $r_0\simeq0.2$ AU (sublimation radius) for Class~0 parameters, $r_0\simeq1$ AU for the Class~I, and $r_0>1$~AU for the Class~II. 
 
\item Key elements for the survival of H$_2$ in MHD disk winds, as opposed
to static irradiated disk atmospheres, are: the short flow crossing
timescales ($50-100$ yrs on the 1 AU streamline), the efficient shielding
provided by inner wind streamlines against stellar FUV and X-ray photons,
and the strong adiabatic cooling which delays gas heating and limits
collisional and chemical destruction.

\item Balance between ion-neutral drag and molecular line cooling establishes asymptotic temperatures $\simeq700/2\,000/3\,000$~K on the 1~AU streamline, increasing with protostellar stage. Temperatures at the wind slow magnetosonic point are within a factor 2 of the surface heating prescribed in the MHD solution, suggesting that the assumed high wind mass-loading might be naturally achievable (although thermal calculations deeper into the disk are needed to test this conjecture).

\item The chemistry of dusty irradiated MHD disk winds is of a complex and hybrid nature, combining the out-of-equilibrium "warm" chemistry of C-shocks heated by ambipolar diffusion, and the photoprocesses present in PDRs and XDRs. Illustrations are the efficient gas-phase synthesis of H$_2$O at all stages, and the substantial abundances reached by SH$^+$, CH$^+$, and $\Hr^+$ in the hotter and more ionized Class~I/II. In these latter cases, OH, O, and C also reach levels similar to H$_2$O while CO is strongly photodissociated.

\item The model predictions appear promising to explain several
observational trends in molecular wind counterparts, namely: the presence in
low-mass Class~0 sources of highly collimated molecular jets with typical
speed up to $100~\kms$, low temperature $\simeq700$~K, and estimated
magneto-centrifugal launch radii $\simeq0.2-0.6$~AU; as well as the trends
for higher H$_2$ temperatures $\simeq2\,500$~K, larger magneto-centrifugal
launch radii, and lower collimation and speed as the source evolves from
Class~0 to Class~I to Class~II.

\item More discriminant tests of the applicability of MHD disk winds to molecular jets/winds in young stars require synthetic maps and line profiles in H$_2$, CO, and H$_2$O, including a non-local treatment of self-shielding and proper account for beam dilution and NLTE excitation. Such calculations will be presented in following papers in this series, and compared with infrared and submm observations. The effect of internal shocks, in particular the enhancement of species believed to be released from grains, such as SiO and CH$_3$OH, will also be explored.
\end{itemize}

\begin{acknowledgements}
We are grateful to G.~Herczeg and P.~Lesaffre for useful suggestions and
discussions, and to the anonymous referee for many insightful comments that
helped to improve the paper. D.~Panoglou, S.~Cabrit, and P.~Garcia wish to
acknowledge financial and travel support through the Marie Curie Research
Training Network JETSET (Jet Simulations, Experiments and Theory) under
contract MRTN-CT-2004-005592. This research has made use of NASA's
Astrophysics Data System.
\end{acknowledgements}

\bibliographystyle{aa}\bibliography{paper-revised}

\appendix

\section{Comparison with the assumed heating in the dynamical solution}
\label{app:heating}
  \begin{figure}
  \centering
  \includegraphics[width=0.35\textwidth]{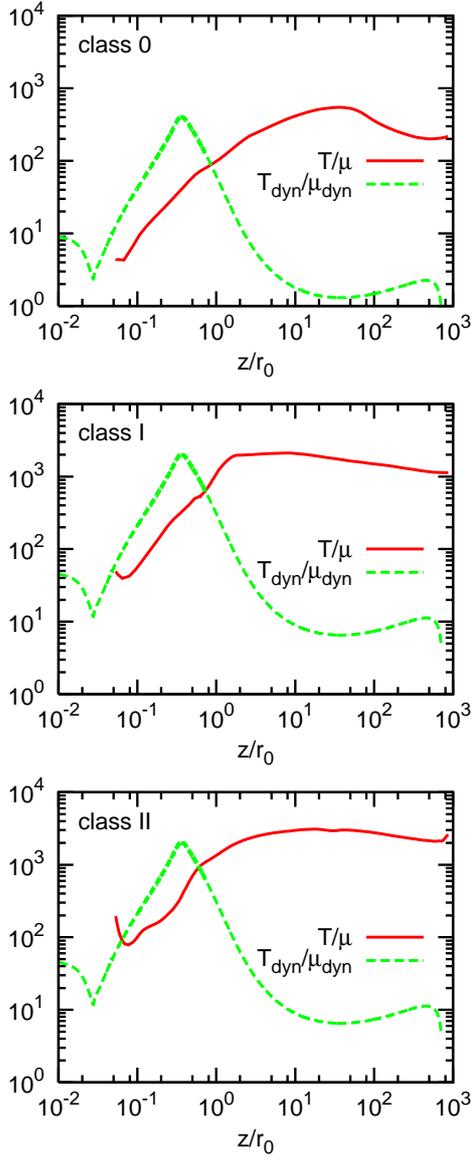}
  \caption{Comparison of $T/\mu$ (in K) as calculated from the
  heating/cooling terms in our thermal-chemical models (red solid curves)
  with the same ratio as assumed in the dynamical MHD solutions of
  \cite{CaFe00}, in dashed green, for the 1 AU streamline and 3 stars of
  differing accretion properties. Note that in the dynamical solution,
  $T_{\rm dyn}$ is calculated assuming a fully ionised gas, with
  $\mu_{\rm dyn}=0.5$. A color version of this figure is available in the
  on-line edition of this journal.}
  \label{f:check} \end{figure}
The MHD solution adopted in the present work has been obtained with a prescribed self-similar enthalpy injection function peaking near the disk surface, representing a tiny fraction $f=8\times10^{-4}$ of the accretion power \citep{CaFe00}. This small surface heating is sufficient to obtain denser, slower disk winds with mass fluxes and poloidal and rotation speeds consistent with observations (see Section~\ref{s:dynamics}), but the nature of the heating process is not yet identified.

In this context, it is interesting to compare the thermal pressure assumed in the underlying MHD accretion-ejection solution with that obtained from our thermal-chemical calculations where the various known heating, cooling and chemical processes are taken into account. Since the gas density $\rho$ is prescribed by the solution, the relevant quantity to compare thermal pressures is just the ratio $T/\mu$ (with $\mu$ being the mean molecular weight). This comparison is shown in Figure~\ref{f:check} along a 1~AU streamline for our Class~0, Class~I, and Class~II models.

The value of $T_{\rm dyn}/\mu_{\rm dyn}$ in the dynamical solution first
undergoes an increase from the disc surface at $z=h=0.03r_0$ to about $z_t=10h$ and then an adiabatic decrease. Although the initial rise is not exactly followed by our calculations, it is interesting that the difference between the two curves is less than a factor 2 near the slow magnetosonic point (our starting integration point for the thermo-chemistry). This suggests that the warm gas surface temperatures required to load dense, slow MHD disk winds compatible with observational constraints might be naturally achievable under a wide range of accretion rates, especially as MHD "turbulent/wave" heating will take place in the deeper resistive disk layers, and wind mass-loading could be further enhanced via upward motions induced by MRI turbulence \citep{Suz09}.

The discrepancy between computed and assumed $T/\mu$ can exceed a factor 10 at $z_t=10h$, and a factor 100 beyond. This discrepancy has however no relevant impact on the dynamical behavior of the solution. Indeed, at $z_t$, the ratio of plasma pressure to magnetic pressure in the dynamical solution is already down to only a few $10^{-3}$, so even a large discrepancy in thermal pressure will not affect the wind dynamics. Although the thermal pressure gradient did play an important role in the mass loading (done within the resistive MHD disk layers), it plays no role in the jet acceleration and collimation itself which is mainly due to magnetic forces. Thermal pressure gradients will eventually start affecting the jet dynamics after the jet refocusses towards the axis. For this reason we have not extended our thermal-chemical calculations beyond the recollimation point of the MHD solution.
\end{document}